\documentclass[notitlepage,aps,prb,onecolumn,groupedaddress]{revtex4-1}

\usepackage{bm}
\usepackage{graphicx}
\usepackage[caption=false,font=footnotesize]{subfig}
\usepackage{booktabs}
\usepackage{textcomp}
\DeclareMathSymbol{\Gamma}{\mathalpha}{letters}{0}
\DeclareMathSymbol{\Theta}{\mathalpha}{letters}{2}
\DeclareMathSymbol{\Lambda}{\mathalpha}{letters}{3}
\DeclareMathSymbol{\Sigma}{\mathalpha}{letters}{6}
\DeclareMathSymbol{\Omega}{\mathalpha}{letters}{10}

\begin{document}

\title{Coupled Nanomechanical Electron Shuttles:\\
Full Stochastic Modeling and Device-Level Simulation}

\author{Mo Zhao}
\affiliation{\mbox{Department of Electrical \rm{\&} Computer Engineering, University of Wisconsin-Madison, Wisconsin 53706, USA}}
\author{Robert H. Blick}
\email{robert@nanomachines.com}
\affiliation{\mbox{Department of Electrical \rm{\&} Computer Engineering, University of Wisconsin-Madison, Wisconsin 53706, USA}}
\affiliation{Center for Hybrid Nanostructures, Falkenried 88, 20251 Hamburg, Germany}
\affiliation{\mbox{Institutes for Nanostructure and Solid State Physics, University of Hamburg, 20355 Hamburg, Germany}}

\begin{abstract}
Earlier theory and measurements show that nanomechanical electron shuttles can work as ratchets for radio-frequency rectification, but its performance was hard to predict so far. This paper focuses on the coupled shuttles which can potentially break symmetry better than a single shuttle.
We propose a full stochastic model of coupled shuttles, where the mechanical motion of nanopillars and the incoherent electronic tunneling are modeled as a Markov chain. A linear master equation is constructed. In particular, the interaction of the their randomness is taken into account. This model favors analyzing the symmetry breaking that results in the observed rectification current~\cite{Kim2010_PRL}. Further, based on the model we propose the deterministic equations of mean physical variables by assuming multivariate Gaussian distribution, which enables complex device simulation and design.
\end{abstract}

\maketitle

\section{Introduction}\label{sec1}
Nanoelectromechanical switches have attracted significant interest in the past decade as they can provide a number of promising applications~\cite{Loh2012_NatNano, Chen2013_ProcIEEE}.
Among them, the nanomechanical electron shuttle proposed by Gorelik~\emph{et al.}~is an outstanding example that received considerable theoretical~\cite{Gorelik1998_PRL, Weiss1999_EPL, Isacsson2004_EPL, Armour2004_69_PRB, Armour2004_70_PRB, Pistolesi2005_PRL, Pistolesi2006_NJP, Huldt2007_NJP, Nishiguchi2008_PRB, Jonsson2008_PRL, Wiersig2008_APL, Pena-Aza2013_PRB, Ahn2006_PRL, Prada2012_PRB} and experimental~\cite{Kim2012_ACSNano, Kim2013, Erbe1998_APL, Park2000_Nature, Erbe2001_PRL, Scheibe2004_APL, Koenig2008_NatNano, Koenig2012_APL, Kim2010_NJP, Kim2010_PRL} attention.
The shuttle is typically realized by nanopillars and excited by a radio-frequency (RF) voltage, leading to electrons being shuttled between two contacts. Due to strong nonlinear electron-mechanical coupling, such devices can be used as RF modulators or as high-frequency current ratchet~\cite{Shekhter2007_JComputTheorNanos}.
Another appealing application of the nanomechanical shuttle lies in energy scavenging. It was shown that single and -- more effectively -- coupled electron shuttles can rectify applied RF signals and give rise to a direct current which can be used to power electronic devices~\cite{Ahn2006_PRL, Prada2012_PRB, Kim2010_PRL, Kim2012_ACSNano, Kim2013}.
While previous theoretical models provide fundamental insight into the physics, they are often limited in accurate predictions on the device level.

In the following paper, we build a new model enabling full analytical and numerical analysis. This is important for designing applications such as a scavenging tool, in which arrays of electron shuttles are coupled to generate an appreciable output current.

\begin{figure}[!htbp]
\centering
\subfloat[]{\includegraphics[height=2cm]{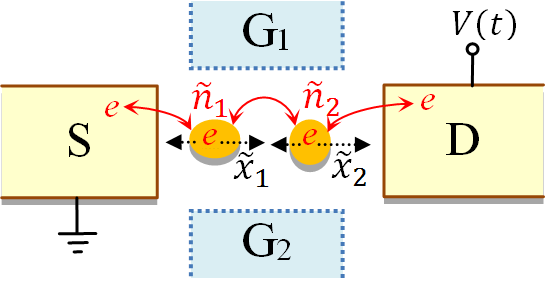}\label{fig1_a}}
\hfil
\subfloat[]{\includegraphics[height=1.7cm]{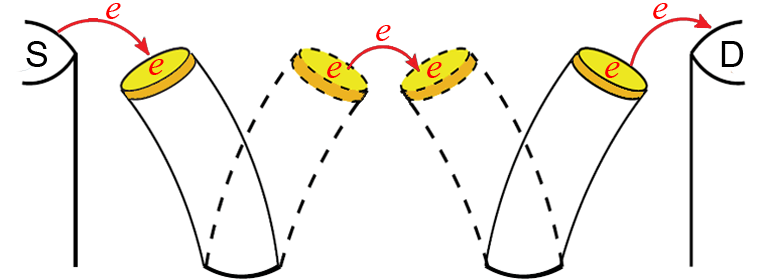}\label{fig1_b}}
\caption{
\label{fig1}
(a) Top-view of the coupled nanomechanical electron shuttles set with the electrodes of source~(S), drain~(D) and optional gates~($\textrm{G}_1$, $\textrm{G}_2$). The two shuttles are slightly different in size or shape, which contain electrons with the number of $\tilde{n}_1(t)$ and $\tilde{n}_2(t)$ and their displacements are denoted by $\tilde{x}_1(t)$ and $\tilde{x}_2(t)$, respectively. An alternating voltage $V(t)$ is applied to the drain electrode, and the source electrode is grounded. (b) Side-view showing vibrations and tunneling of the coupled electron shuttles that are on top of nanopillars.}
\end{figure}

As illustrated in Fig. \ref{fig1}, the coupled electron shuttles oscillate between source and drain electrodes under voltage $V(t)$. Optionally, we may add the gate electrodes biased with a constant charge (electron number noted by vector $\bm n_{\rm G}$).
Because electron tunneling is random and incoherent for different junctions, we model tunneling and the mechanical motion in the system as a continuous stochastic process.
Hence, a semi-classical statistical model is better suited than the full quantum-mechanical treatment~\cite{Fedorets2004_PRL,Novotny2003_PRL,Novotny2004_PRL}.
We develop a linear master equation describing the probability distribution of the electron numbers in the shuttles.
Although the probability distribution was widely discussed in previous studies on the single shuttle~\cite{Gorelik1998_PRL, Weiss1999_EPL, Pistolesi2005_PRL, Pistolesi2006_NJP, Huldt2007_NJP, Nishiguchi2008_PRB, Jonsson2008_PRL, Wiersig2008_APL, Pena-Aza2013_PRB}, most approaches describe the mechanical vibrations by deterministic variables and hence derive a master equation being nonlinear.

Ahn~\cite{Ahn2006_PRL} and Prada~\cite{Prada2012_PRB} extended such method to pioneer modeling the coupled shuttles. In contrast to some earlier models~\cite{Armour2004_69_PRB, Armour2004_70_PRB, Pistolesi2005_PRL, Pistolesi2006_NJP, Huldt2007_NJP, Nishiguchi2008_PRB, Pena-Aza2013_PRB}, they allow for electron numbers in a shuttle to be greater than 1. Our model follows their assumptions but enables further analytical and numerical work. More importantly, we propose the deterministic equations for the mean value of properties including the electron current. This brings the large-scale device simulation to a level of acceptable speed and accuracy.

\section{Full Stochastic Model}\label{sec2}
We describe the displacement and velocity of the $s$\textsuperscript{th} shuttle ($s=1,2$) at the time $t$ by random variables $\tilde{x}_s (t)$ and $\tilde{v}_s (t)$, and describe the number of net electrons (with charge noted by $-q$ for each) in the shuttle by the integer-valued random variable $\tilde{n}_s (t)$. For simplicity, we write them by vectors $\tilde{\bm x}(t)$, $\tilde{\bm v}(t)$ and $\tilde{\bm n}(t)$ (typically $(t)$ is omitted in the following). We assume these three random variables are sufficient to describe the immediate state of the system and evolute as a Markov chain. The mean of a random variable is noted by a bracket, e.g., $\langle\tilde{\bm x}(t) \rangle$ or simply $\langle\tilde{\bm x} \rangle$. We use $P(\bm n,\bm x,\bm v,t)$ to describe the joint probability distribution function (PDF) of $\tilde{\bm n}(t)$, $\tilde{\bm x}(t)$ and $\tilde{\bm v}(t)$, whose variables $\bm n$, $\bm x$ and $\bm v$ have the same value range as these random variables. Assume after an infinitesimal time $\Delta t = t' - t$, the PDF $P(\bm n,\bm x,\bm v,t)$ changes to $P(\bm n',\bm x',\bm v',t')$, and we define $\Delta x_s = x'_s - x_s$ and $\Delta v_s = v'_s - v_s$. The mechanical vibration of nanopillars and the tunneling of electrons are two correlated mechanisms governing the dynamics of the system. Let us discuss them separately in the following context.

The fundamental mode of the nanopillar vibration is the most interesting for us, since it has the greatest amplitude among all modes for the same driven energy. It can be modeled by the one-dimensional vibration characterized by eigenfrequency, which depends on the size and material of the nanopillars.
Assuming the $s$\textsuperscript{th} pillar has an eigenfrequency ${\omega}_s$, effective mass $m_s$, and damping coefficient ${\gamma}_s = {\omega}_s/Q$ with $Q$ being the quality factor, we have $\tilde{x}_s(t)$ and $\tilde{v}_s(t)$ satisfying the following stochastic differential equations:

\begin{equation}\label{eq_1}
\left\{ \begin{array}{l}
{\rm d} \tilde{x}_s / {\rm d} t = \tilde{v}_s \\
{\rm d} \tilde{v}_s / {\rm d} t = -\gamma_s \tilde{v}_s - \omega_s^2 \tilde{x}_s + F_s(\tilde{\bm n},t)/m_s
\end{array} \right.
\end{equation}

\noindent where $F_s$ is the electromagnetic force on the $s$th shuttle and can be approximated as a function of $\tilde{\bm n}(t)$ and $V(t)$ (discussed in the next section). Solving this equation, we can describe $\tilde{x}_s$ and $\tilde{x}_s$ in terms of the Ito integral with $F_s (\tilde{\bm n},t)$. By discarding the transient-state part of solution which damps over time by a factor of $e^{-(\gamma_s t)/2}$, we have the steady-state solution:
\begin{equation}\label{eq_0a}
\tilde{x}_s (t)=\left[\sin(\omega'_s t) \tilde{J}_s^{\rm{cos}}(t)- \cos(\omega'_s t) \tilde{J}_s^{\rm{sin}}(t)\right]/(\omega'_s m_s ),
\end{equation}
\begin{equation}\label{eq_0b}
\tilde{v}_s (t)=\left[\cos(\omega'_s t+\phi_s ) \tilde{J}_s^{\rm{cos}} (t)+\sin(\omega'_s t+\phi_s ) \tilde{J}_s^{\rm{sin}}(t)\right] (\omega_s/\omega'_s )/m_s,
\end{equation}

\noindent where $\omega'_s=\left[\omega_s^2-(\gamma_s^2)/4\right]^{1/2}$, $\phi_s = \tan^{-1}(\gamma_s/{2\omega'_s})$, and
\begin{equation}\label{eq_0c}
\tilde{J}_s^{\rm{sin}} (t)=e^{-(\gamma_s t)/2} \int_0^t F_s (\tilde{\bm n},\tau) e^{(\gamma_s t)/2}  \sin(\omega'_s \tau)\rm{d}t,
\end{equation}
\begin{equation}\label{eq_0d}
\tilde{J}_s^{\rm{cos}} (t)=e^{-(\gamma_s t)/2} \int_0^t F_s (\tilde{\bm n},\tau) e^{(\gamma_s t)/2}  \cos(\omega'_s \tau)\rm{d}t.
\end{equation}
\noindent However, we notice that this form of solution only makes sense when the probability distribution of $\tilde{\bm n}(t)$ is known. Instead, we try to formulate the conditional probability $P(x'_s,v'_s,t'|\bm n,\bm x,\bm v,t)$. For the events in the sample space tagged with $\tilde{\bm x}(t)=\bm x$ and $\tilde{\bm v}(t)=\bm v$, the stochastic differential equation (\ref{eq_1}) can predict the increment of $\tilde{\bm x}(t)$ and $\tilde{\bm v}(t)$ at time $t$, i.e., after $\Delta t \rightarrow 0$,
\begin{equation}\label{eq_1b}
\left\{ \begin{array}{l}
\tilde{x}_s(t'|\bm n,\bm x,\bm v) = x_s + v_s \Delta t + \tilde{o}(\Delta t)\\
\tilde{v}_s(t'|\bm n,\bm x,\bm v) = v_s + (-\gamma_s v_s - \omega_s^2 x_s + F_s(\bm n,t)/m_s) \Delta t + \tilde{o}(\Delta t)
\end{array} \right.
\end{equation}

\noindent where $\tilde{o}(\Delta t)$ denotes a random variable being a higher-order infinitesimal of $\Delta t$. Thus, for the first order, the displacement and velocity at time $t'$ is determinate, and the conditional probability should be a Dirac delta function:
\begin{equation}\label{eq_2}
P(x'_s,v'_s,t'|\bm n,\bm x,\bm v,t)
= \delta \left(x'_s-x_s - v_s \Delta t \right)
\delta \left(v'_s-v_s + [\gamma_s v_s + \omega_s^2 x_s - F_s(\bm n,t)/m_s] \Delta t\right)
+ o(\Delta t),
\end{equation}

\noindent where $\delta(x'_s - \ldots)$ denotes the Dirac delta function.

This conclusion can be strictly proved. We can follow the principle of Kramers-Moyal expansion~\cite{mahnke2009physics} to account for the evolution of $\tilde{\bm x}(t)$ and $\tilde{\bm v}(t)$, although the expansion fails for $P(\bm n,\bm x,\bm v,t)$ because the change of $\tilde{\bm n}(t)$ has special properties and should be separately considered. We may start from the $l_1,l_2$-order mixed derivative moment of $\tilde{x}_s$ and $\tilde{v}_s$ defined by
\begin{eqnarray}\label{eq_B0a}
\xi_s^{(l_1 l_2)}(\bm n,\bm x,\bm v,t)
&\equiv& \lim_{t'\rightarrow t}\frac{1}{t'-t} \int \!\!\! \int
(x'_s-x_s) ^{l_1} (v'_s-v_s)^{l_2} P(x'_s,v'_s,t'|\bm n,\bm x,\bm v,t)
{\rm d}x'_s {\rm d}v'_s,
\end{eqnarray}

\noindent where the region of integration is the real plane ${\bf R}^2$, and $l_1$ and $l_2$ are positive integers. This integral denotes a conditional mean. For formulation, we can expand $\tilde{x}_s(t')$ and $\tilde{v}_s(t')$ by Taylor series to express the increments and ignore infinitesimals with the order higher than $\Delta t$ in the denominator. Using (\ref{eq_1}), we can derive
\begin{eqnarray}\label{eq_B0b}
\xi_s^{(l_1 l_2)}(\bm n,\bm x,\bm v,t)
= \left\{ \begin{array}{ll}
v_s, & l_1=1,l_2=0, \\
-\gamma_s v_s - \omega_s^2 x_s + F_s(\bm n,t)/m_s, & l_1=0,l_2=1, \\
0, & l_1 + l_2 \geq 2.
\end{array} \right.
\end{eqnarray}

\noindent This can be used to truncate the expansion series of $P(x'_s,v'_s,t'|\bm n,\bm x,\bm v,t)$ so that $l_1 + l_2 \leq 1$. Hence, we define the Fourier transform of $P(x'_s,v'_s,t'|\bm n,\bm x,\bm v,t)$ by
\begin{eqnarray}\label{eq_B1a}
\bar{P}(k_s,w_s,t'|\bm n,\bm x,\bm v,t)
\equiv \int \!\!\! \int
{\rm d}x'_s {\rm d}v'_s e^{i(k_s x'_s + w_s v'_s)}
P(x'_s,v'_s,t'|\bm n,\bm x,\bm v,t),
\end{eqnarray}

\noindent which, with a factor of $e^{-i(k_s x_s + w_s v_s)}$, can be expanded by the Taylor series:
\begin{eqnarray}\label{eq_B1b}
\bar{P}(k_s,w_s,t'|\bm n,\bm x,\bm v,t) e^{-i(k_s x_s + w_s v_s)}
= \sum_{l_1,l_2=0}^{\infty}
\frac{i^{l_1+l_2}}{l_1!\,l_2!}
k_s^{l_1} w_s^{l_2}
\int \!\!\! \int
{\rm d}x'_s {\rm d}v'_s
(x'_s-x_s)^{l_1} (v'_s-v_s)^{l_2}
P(x'_s,v'_s,t'|\bm n,\bm x,\bm v,t).
\end{eqnarray}
With (\ref{eq_B0a}), it becomes
\begin{eqnarray}\label{eq_B1c}
\bar{P}(k_s,w_s,t'|\bm n,\bm x,\bm v,t)
= e^{i(k_s x_s + w_s v_s)}
\left[1 + \Delta t \sum_{l_1+l_2=1}
\frac{1}{l_1!\,l_2!}
\xi_s^{(l_1 l_2)}(\bm n,\bm x,\bm v,t)
(i k_s)^{l_1} (i w_s)^{l_2}
+ o(\Delta t) \right],
\end{eqnarray}

\noindent where $o(\Delta t)$ denotes the higher-order infinitesimal of $\Delta t$. Consider the inverse Fourier transform:
\begin{eqnarray}\label{eq_B1d}
P(x'_s,v'_s,t'|\bm n,\bm x,\bm v,t)
= \frac{1}{2 \pi} \int \!\!\! \int
{\rm d}k_s {\rm d}w_s e^{-i(k_s x'_s + w_s v'_s)}
\bar{P}(k_s,w_s,t'|\bm n,\bm x,\bm v,t).
\end{eqnarray}

\noindent Substituting (\ref{eq_B1c}) in (\ref{eq_B1d}) and change the sequence of integral and summation, we obtain
\begin{eqnarray}\label{eq_B1e}
P(x'_s,v'_s,t'|\bm n,\bm x,\bm v,t)
&=& \frac{1}{2 \pi} \int \!\!\! \int {\rm d}k_s {\rm d}w_s
e^{i[k_s (x_s-x'_s) + w_s (v_s-v'_s)]}\\
&+& \Delta t \sum_{l_1+l_2=1}
\frac{1}{l_1!\,l_2!}
\xi_s^{(l_1 l_2)}(\bm n,\bm x,\bm v,t)
\frac{1}{2 \pi} \int \!\!\! \int
{\rm d}k_s {\rm d}w_s
(i k_s)^{l_1} (i w_s)^{l_2}
e^{i[k_s (x_s-x'_s) + w_s (v_s-v'_s)]}
+o(\Delta t)\nonumber\\
&=& \delta(x'_s-x_s)\delta(v'_s-v_s)
+\Delta t \!\! \sum_{l_1+l_2=1}
\frac{(-1)^{l_1+l_2}}{l_1!\,l_2!}
\xi_s^{(l_1 l_2)}(\bm n,\bm x,\bm v,t)
\frac{\partial^{l_1}\delta(x'_s-x_s)}{(\partial x'_s)^{l_1}}
\frac{\partial^{l_2}\delta(v'_s-v_s)}{(\partial v'_s)^{l_2}}
+o(\Delta t).\nonumber
\end{eqnarray}

\noindent With (\ref{eq_B0b}) substituted in, it can be written explicitly as
\begin{eqnarray}\label{eq_B1f}
P(x'_s,v'_s,t'|\bm n,\bm x,\bm v,t)
&=& \delta(x'_s-x_s)\delta(v'_s-v_s) \\
&-& \Delta t
\left\{\delta'(x'_s-x_s)\delta(v'_s-v_s) v_s
+\delta(x'_s-x_s)\delta'(v'_s-v_s) \left[-\gamma_s v_s - \omega_s^2 x_s + F_s(\bm n,t)/m_s \right] \right\}
+o(\Delta t).\nonumber\
\end{eqnarray}
where $\delta'(x'_s-x_s)$ denotes the derivative of the Dirac delta function. This is obviously the Taylor expansion of (\ref{eq_2}).

Another mechanism restraining the PDF besides the mechanical motion is the tunneling of electrons, which can be modeled by the orthodox model \cite{Armour2004_69_PRB}. The components of $\bm n- \bm n'$ are limited to $\pm$1 and 0 for considering single electron tunneling at each step. Technically, we define a vector $\bm\mu$, whose components $\mu_j$ denote the number of electron tunneled through the $j$\textsuperscript{th} junction ($j=1,2,3$) and can be 0 and $\pm$1. We also define
\begin{equation}\label{eq_3}
\bm T \equiv
\left[ \begin{array}{rr}
1 & 0 \\
-1 & 1 \\
0 & -1
\end{array} \right]
\end{equation}
\noindent so that $\bm n=\bm n' - \bm\mu \bm T$ and note its $j$\textsuperscript{th} row by a vector $\bm T_j$. The transition probability of $\mu_j$ electrons tunneling through the $j$\textsuperscript{th} junction is
\begin{eqnarray}\label{eq_6}
P(\mu_j,t'|\bm n,\bm x, t) = \left\{ \begin{array}{ll}
\Gamma_j^{\pm}(\bm n,\bm x,t) \Delta t, & \mu_j=\pm 1, \\
1 - [\Gamma_j^{+}(\bm n,\bm x,t) + \Gamma_j^{-}(\bm n,\bm x,t)] \Delta t, & \mu_j=0,
\end{array} \right.
\end{eqnarray}

\noindent where
\begin{equation}\label{eq_7}
\Gamma_j^{\pm}(\bm n,\bm x,t) = \frac{{\rm e}^{-\bm x \cdot \bm T_j/\lambda_j}}{q^2R_j^0}\frac{U_j^{\pm}(\bm n,\bm x,t)}{1 - {\rm e}^{-U_j^{\pm}(\bm n,\bm x,t)/k_{\rm B}T}}
\end{equation}
\noindent is the forward/backward ($+/-$) tunneling rate of electrons through the $j$\textsuperscript{th} junction, wherein $R_j^0$ is the unperturbed tunneling resistance from the mechanical motions, $\lambda_j$ is the tunneling wavelength, $T$ is the temperature, $k_{\rm B}$ is the Boltzmann constant, and $U_j^{\pm}$ denotes the change of electromagnetic energy due to the tunneling.

\noindent is the forward/backward ($+/-$) tunneling rate of electrons through the $j$\textsuperscript{th} junction, wherein $R_j^0$ is the unperturbed tunneling resistance from the mechanical motions, $\lambda_j$ is the tunneling wavelength, $T$ is the temperature, $k_{\rm B}$ is the Boltzmann constant. $U_j^{\pm}(\bm n,\bm x,t)$ denotes the change of electromagnetic energy before and after this electron tunneling. It is a function of $\bm n$, $\bm x$ and $V(t)$, and is elaborated in the next section. In general cases, $U_j^{\pm}(\bm n,\bm x,t)$ loosely depends on $\bm x$ and we can separate $\bm n$ and $\bm x$ so that $\Gamma_j^{\pm}(\bm n,\bm x,t) = K_j({\bm x}) \Gamma_j^{\pm}(\bm n,t)$ by defining
\begin{equation}\label{eq_B6}
K_j({\bm x})={\rm e}^{-{\bm x} \cdot \bm T_j/\lambda_j}
\end{equation}
and
\begin{equation}\label{eq_B7}
\Gamma_j^{\pm}(\bm n,t) = \frac{1}{q^2R_j^0}\frac{U_j^{\pm}(\bm n,t)}{1 - {\rm e}^{-U_j^{\pm}(\bm n,t)/k_{\rm B}T}}
\end{equation}

Because the electron tunneling through each junction and the mechanical motion of each shuttle are independent, we can write the Chapman-Kolmogorov equation~\cite{mahnke2009physics}:
\begin{eqnarray}\label{eq_4}
P(\bm n',\bm x',\bm v',t') =
\int{\rm d}\Omega \! \sum_{\mu_j=0,\pm 1} \prod_{j=1}^{3}{P(\mu_j,t'|\bm n,\bm x,t)} \prod_{s=1}^{2}{P(x'_s, v'_s, t'|\bm n,\bm x,\bm v,t)} P(\bm n,\bm x,\bm v,t)
\end{eqnarray}

\noindent where ${\rm d}\Omega={\rm d}x_1 {\rm d}x_2 {\rm d}v_1 {\rm d}v_2$. Substitute (\ref{eq_6}) into (\ref{eq_4}), and only count the terms that are the same order infinitesimal of $\Delta t$. Then, the integrand of (\ref{eq_4}) becomes
\begin{eqnarray}\label{eq_B3}
&&\left[1 - \Delta t \sum_{j=1}^3\sum_{\pm}\!\!\Gamma_j^{\pm}(\bm n',\bm x,t)\right]
\prod_{s=1}^{2}{P(x'_s, v'_s, t'|\bm n',x_s,v_s,t)}P(\bm n',\bm x,\bm v,t) \nonumber\\
&&+ \Delta t \sum_{j=1}^3\sum_{\pm}\!\!\Gamma_j^{\pm}(\bm n' \mp \bm T_j,\bm x,t)
\prod_{s=1}^{2}{P(x'_s, v'_s, t'|\bm n' \mp \bm T_j,x_s,v_s,t)}P(\bm n' \mp \bm T_j,\bm x,\bm v,t)]
+ o(\Delta t).
\end{eqnarray}

\noindent With (\ref{eq_B1f}), (\ref{eq_B3}) becomes
\begin{eqnarray}\label{eq_B4}
&&\left[1 - \Delta t \sum_{j=1}^3\sum_{\pm}\!\!\Gamma_j^{\pm}(\bm n',\bm x,t)\right]
\delta (\bm v - \bm v')\delta (\bm x - \bm x') P(\bm n',\bm x',\bm v',t) \nonumber\\
&&- \Delta t P(\bm n',\bm x',\bm v',t)
\sum_{s=1}^{2} \left[v_s \frac{\partial}{\partial x'_s}
+\left(-\gamma_s v_s - \omega_s^2 x_s + \frac{F_s(\bm n,t)}{m_s} \right) \frac{\partial}{\partial v'_s} \right]
\prod_{s=1}^{2} \delta(x'_s-x_s)\delta'(v'_s-v_s) \nonumber\\
&&+ \Delta t \sum_{j=1}^3\sum_{\pm}\!\!\Gamma_j^{\pm}(\bm n' \mp \bm T_j,\bm x,t)
\delta (\bm v - \bm v')\delta (\bm x - \bm x') P(\bm n' \mp \bm T_j,\bm x',\bm v',t)
+ o(\Delta t).
\end{eqnarray}

\noindent The integral in (\ref{eq_4}) can be calculated with the property of the delta function:
\begin{eqnarray}\label{eq_B5}
P(\bm n',\bm x',\bm v',t')
&&= P(\bm n',\bm x',\bm v',t)
+ \Delta t \sum_{j=1}^3\sum_{\pm }\!
\left[ \Gamma_j^{\pm}(\bm n' \mp \bm T_j,\bm x,t)
P(\bm n' \mp \bm T_j,\bm x',\bm v',t)
-\Gamma_j^{\pm}(\bm n',\bm x,t)
P(\bm n',\bm x',\bm v',t) \right] \nonumber\\
&&+ \Delta t \sum_{s=1}^{2} \left[v_s \frac{\partial}{\partial x_s}
- \frac{\partial}{\partial v_s} \left(-\gamma_s v_s - \omega_s^2 x_s + \frac{F_s(\bm n,t)}{m_s}\right) \right]
P(\bm n',\bm x',\bm v',t)
+ o(\Delta t).
\end{eqnarray}

\noindent By moving $P(\bm n',\bm x',\bm v',t)$ to the left side and dividing both sides by $\Delta t$, we can formulate the following master equation
\begin{eqnarray}\label{eq_B11}
&&\frac{\partial P(\bm n,\bm x,\bm v,t)}{\partial t}
= \sum_{j=1}^{3}\sum_{\pm}[\Gamma_j^\pm(\bm n\mp\bm T_j,\bm x,t)
P(\bm n\mp\bm T_j,\bm x,\bm v,t)
- \Gamma_j^\pm (\bm n,\bm x,t) P(\bm n,\bm x,\bm v,t)] \nonumber\\
&&+\sum_{s = 1}^{2} \left[ \gamma_s P(\bm n,\bm x,\bm v,t)
- v_s \frac{\partial P(\bm n,\bm x,\bm v,t)}{\partial x_s}
+ \left(\gamma_s v_s+\omega_s^2 x_s-\frac{F_s(\bm n,t)}{m_s}\right)
\frac{\partial P(\bm n,\bm x,\bm v,t)}{\partial v_s} \right].
\end{eqnarray}

\noindent It can also be simply noted by
\begin{eqnarray}\label{eq_9}
\frac{\partial P}{\partial t}= \sum_{j = 1}^{3}\sum_{\pm}\left(\hat{\mathcal{N}}_{\mp\bm T_j} - 1\right) \Gamma_j^{\pm} P
+\sum_{s = 1}^{2}\left[\gamma_s P - v_s \frac{\partial P}{\partial x_s}
+ \left(\gamma_s v_s+\omega_s^2 x_s-\frac{F_s(\bm n,t)}{m_s} \right)
\frac{\partial P}{\partial v_s}\right].
\end{eqnarray}

\noindent where the operator $\hat{\mathcal{N}}_{\mp\bm T_j}$ shifts the argument $\bm n$ in a function by $\mp\bm T_j$. Eq~(\ref{eq_9}) is a linear first-order partial differential equation. Its boundary condition is implicit, i.e., as $|n_s|\rightarrow\infty$ or $|x_s|\rightarrow\infty$ or $|v_s|\rightarrow\infty$, $P(\bm n,\bm x,\bm v,t)\rightarrow 0$ (asymptotically as a Gaussian function). In addition, (\ref{eq_9}) is homogeneous, so the solution is linear with the initial condition, in which the PDF should be normalized. Given an initial condition, we can solve the equation to obtain a conditional-PDF.

Nevertheless, we are more interested in the steady-state solution in which the PDF becomes periodic and irrelevant to the initial PDF after numerous periods (so that the initial condition only plays a role in normalization). Such solution is possible because the coefficients of the equation are periodic in time and the influence of the initial condition dies out. In fact, the periodicity and the irrelevance with the initial condition are corresponding. After the period $2\pi/\omega$, the coefficients of (\ref{eq_9}) remain the same while the unknown function is changed from $P(\bm n,\bm x,\bm v,t)$ to $P(\bm n,\bm x,\bm v,t+2\pi/\omega)$. The two functions should be the same if there is a unique steady-state solution irrelevant to the initial condition. If the function is periodic, due to the linear property of the differential equation, the combination of solutions with arbitrary weights remains, but the initial condition could be varied.

\begin{figure}[!t]
\centerline{\includegraphics[width=13cm]{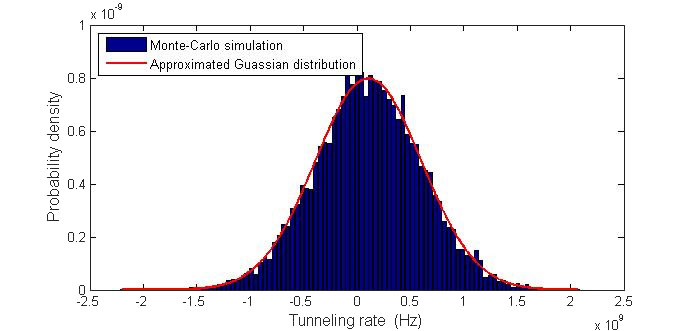}}
\caption{\label{fig2}Probability distribution for the tunneling rate obtained from the Monte-Carlo simulation as well as the deterministic method with the assumption of multivariate Gaussian distribution.}
\end{figure}

From numerical solutions of (\ref{eq_9}), we learn that such a steady-state PDF (at a specific time) is usually very close to a multivariate Gaussian distribution, as a typical example shown in Fig. \ref{fig2}. This conclusion complies with the central limit theorem under weak dependence, because the stochastic process consists of infinite times of single electron tunneling which are weakly dependent on each other.

\section{Electromagnetic Energy and Force}\label{sec4}

In principle, the electromagnetic field can be modeled by the partial capacitance matrix due to the geometry that makes the inductance (or variation of the magnetic energy) negligible (this can be seen from the frequency sweep of inductance from electromagnetic simulation). Thus, the energy ${E}_{\rm C}$ stored in the field should be a homogeneous 2\textsuperscript{nd}-order polynomial of the charges $q{\bm n}(t)$, $q\bm n_{\rm G}$ and the applied voltage $V(t)$, with coefficients associated with the partial capacitance.

We denote the charges stored on the shuttles by a vector $\bm Q_{\rm S}$. Because the size of shuttles is small and they are made of metals with high conductivity, we can ignore the variation of electric potentials in the shuttles and denote their potential by $\bm V_{\rm S}$ with the source electrode grounded. Similarly, we use $\bm Q_{\rm G}$ and $\bm V_{\rm G}$ to denote the charge and potential of the biased gate, and use $Q_0$ and $V$ for the right electrode. For convenience, we define $\bm Q_{\rm{all}} \equiv [\bm Q_{\rm S}, \bm Q_{\rm G}, Q_0]$, $\bm V_{\rm{all}} \equiv [\bm V_{\rm S}, \bm V_{\rm G}, V]$, and $\bm Q_{\rm{eff}} \equiv [\bm Q_{\rm{S}}, \bm Q_{\rm{G}}]$, $\bm V_{\rm{eff}} \equiv [\bm V_{\rm{S}}, \bm V_{\rm{G}}]$. The dynamic electromagnetic system satisfies the Laplace's equation: we assume the charges are distributed on the boundary of metal islands, and linearity of the equation make potential distribution proportional to the charge quantity in the boundary condition, i.e., $\bm V_{\rm{all}} = \bm Q_{\rm{all}}\bm C_{\rm{all}}^{-1}$, wherein the coefficient matrix is the capacitance matrix $\bm C_{\rm{all}}$ which can be defined by a form of block matrix:
\begin{equation}\label{eq_A1}
\bm C_{\rm{all}} \equiv
\left[ \begin{array}{cc}
\bm C & \bm c^{\rm{T}} \\
\bm c & C_{00}
\end{array} \right]
\end{equation}

\noindent where $\bm c = [\bm c_{\rm{S}}, \bm c_{\rm{G}}]$ has the same dimension as $\bm Q_{\rm{eff}}$, $C_{00}$ is a scalar, and
\begin{equation}\label{eq_A2}
\bm C \equiv
\left[ \begin{array}{cc}
\bm C_{\rm{SS}} & \bm c_{\rm{GS}}^{\rm{T}} \\
\bm c_{\rm{GS}} & \bm C_{\rm{GG}}
\end{array} \right]
\end{equation}

We should first describe the electromagnetic energy $E_{\rm C}$ in terms of $\bm Q_{\rm{eff}}$ and $V$. From (\ref{eq_A1}) we get
\begin{equation}\label{eq_A3}
\bm Q_{\rm{eff}} = \bm V_{\rm{eff}} \bm C + V \bm c
\end{equation}
\begin{equation}\label{eq_A4}
Q_0 = \bm V_{\rm{eff}} \bm c^{\rm{T}} + VC_{00}
\end{equation}

\noindent Eq (\ref{eq_A3}) can be rewritten as
\begin{equation}\label{eq_A5}
\bm V_{\rm{eff}} = (\bm Q_{\rm{eff}} - V \bm c) \bm C^{-1}
\end{equation}

\noindent where the inverse of $\bm C$ has the following explicit form:
\begin{equation}\label{eq_A6}
\bm C^{-1} =
\left[ \begin{array}{cc}
\bm C_{\rm{S}}^{-1} & -\bm C_{\rm S}^{-1}\bm B^{\rm{T}}\\
-\bm B \bm C_{\rm{S}}^{-1} & \bm C_{\rm{G}}^{-1}
\end{array} \right]
\end{equation}

\noindent where
\begin{equation}\label{eq_A7}
\bm B = \bm C_{\rm{GG}}^{-1} \bm c_{\rm{GS}}
\end{equation}
\begin{equation}\label{eq_A8}
\bm C_{\rm{S}}^{-1} = \bm C_{\rm{SS}}^{-1} + \bm C_{\rm{SS}}^{-1} \bm c_{\rm{GS}}^{\rm{T}}\bm C_{\rm{G}}^{-1} \bm c_{\rm{GS}} \bm C_{\rm{SS}}^{-1}
\end{equation}
\begin{equation}\label{eq_A9}
\bm C_{\rm G}^{-1} = \bm C_{\rm{GG}}^{-1} + \bm B \bm C_{\rm{S}}^{-1} \bm B^{\rm{T}}
\end{equation}

\noindent Substituting (\ref{eq_A5}) and (\ref{eq_A6}) into the (\ref{eq_A4}) we obtain
\begin{equation}\label{eq_A10}
Q_0 = \bm Q_{\rm{eff}} \bm C^{-1} \bm c^{\rm{T}} + C_0 V = (\bm Q_{\rm S} - \bm Q_{\rm G}\bm B)\bm\zeta^{\rm T} + \bm Q_{\rm G}\bm C_{\rm{GG}}^{-1}\bm c_{\rm G}^{\rm{T}}  + C_0 V,
\end{equation}

\noindent where
\begin{equation}\label{eq_A11}
C_0 = C_{00} - \bm c \bm C^{-1} \bm c^{\rm{T}},
\end{equation}
\begin{equation}\label{eq_A12}
\bm\zeta = (\bm c_{\rm S} - \bm c_{\rm G}\bm B) \bm C_{\rm{S}}^{-1}.
\end{equation}

\noindent Note that $C_0$ is the capacitance seen from the two electrodes if no charges on shuttles, and $\bm C_{\rm G}$ is the capacitance if the test voltage is on the gate terminal and charges on shuttles are ignored. These values hardly change with the movement of shuttles.

If we ignore the potential variation in a metal island as well as the magnetic potential which is very small for our geometry, the energy of electromagnetic field can be well approximated by
\begin{equation}\label{eq_A13}
E_{\rm C} = \frac{1}{2} \bm V_{\rm{all}} \bm Q_{\rm{all}}^{\rm{T}}.
\end{equation}

\noindent Substituting (\ref{eq_A5}) and (\ref{eq_A10}) into (\ref{eq_A13}), we obtain
\begin{eqnarray}\label{eq_A14}
E_{\rm C} &=& \frac{1}{2} (\bm Q_{\rm{eff}} - V \bm c) \bm C^{-1} \bm Q_{\rm{eff}}^{\rm{T}} + \frac{1}{2} V(\bm Q_{\rm{eff}} \bm C^{-1} \bm c^{\rm{T}} + C_0 V) \nonumber\\
&=& \frac{1}{2} \bm Q_{\rm{eff}} \bm C^{-1} \bm Q_{\rm{eff}}^{\rm{T}} + \frac{1}{2} C_0 V^2.
\end{eqnarray}

\noindent Substituting (\ref{eq_A6}) and (\ref{eq_A9}) into (\ref{eq_A14}), we get
\begin{eqnarray}\label{eq_A15}
E_{\rm C} &=& \frac{1}{2} \bm Q_{\rm S} \bm C_{\rm S}^{-1} \bm Q_{\rm S}^{\rm{T}} - \bm Q_{\rm S} \bm C_{\rm{S}}^{-1} \bm B^{\rm{T}}\bm Q_{\rm G}^{\rm{T}} + \frac{1}{2}\bm Q_{\rm G}(\bm C_{\rm{GG}}^{-1} + \bm B\bm C_{\rm{S}}^{-1}\bm B^{\rm T})\bm Q_{\rm G}^{\rm{T}} + \frac{1}{2}C_0 V^2 \nonumber\\
&=&\frac{1}{2}(\bm Q_{\rm S} - \bm Q_{\rm G}\bm B)\bm C_{\rm{S}}^{-1}(\bm Q_{\rm S} - \bm Q_{\rm G}\bm B)^{\rm T} + \frac{1}{2}\bm Q_{\rm G}\bm C_{\rm{GG}}^{-1}\bm Q_{\rm G}^{\rm{T}} + \frac{1}{2}C_0 V^2.
\end{eqnarray}

The electromagnetic force is a conservative force, which is the spatial derivative of the stored energy, so we can choose an arbitrary path to calculate the force. If the charge $Q_0$ on electrode is set as constant, $V$ is related to $Q_0$ by
\begin{equation}\label{eq_A16}
V =C_0^{-1}[-(\bm Q_{\rm S} - \bm Q_{\rm G}\bm B)\bm\zeta^{\rm T} - \bm Q_{\rm G} \bm C_{\rm GG}^{-1}\bm c_{\rm G}^{\rm T}+Q_0],
\end{equation}

\noindent which is simply converted from (\ref{eq_A10}). Thus, the corresponding force on the $s$\textsuperscript{th} shuttle ($s=1,2$) can be derived from $F_s = -\partial E_{\rm C}/\partial x_s$, and it can be written as a function of the electron number $\tilde{\bm n} = -\bm Q_{\rm S}/q$ and the time:
\begin{equation} \label{eq_A17}
F_s({\bm n},t) = {\bm n} \bm F_s^0 {\bm n}^{\rm T} + {\bm n} \bm F_s^{\rm G} \bm n_{\rm G}^{\rm T} + \bm n_{\rm G} \bm F_s^{\rm GG} \bm n_{\rm G}^{\rm T}
q({\bm n} \cdot \bm{\alpha}_s + \bm n_{\rm G} \cdot \bm\alpha_s^{\rm G})V(t),
\end{equation}

\noindent where we define
\begin{equation}\label{eq_A18}
\bm F_s^0 \equiv -\frac{1}{2}q^2\partial \bm C_{\rm S}^{-1}/\partial x_s,
\end{equation}
\begin{equation}\label{eq_A19}
\bm F_s^{\rm G} \equiv q^2\partial (\bm C_{\rm S}^{-1}\bm B^{\rm T})/\partial x_s,
\end{equation}
\begin{equation}\label{eq_A20}
\bm \alpha_s \equiv -\partial \bm \zeta/\partial x_s,
\end{equation}
\begin{equation}\label{eq_A21}
\bm \alpha_s^{\rm G} \equiv \partial(\bm\zeta\bm B^{\rm T} - \bm c_{\rm{G}}\bm C_{\rm{GG}}^{-1})/\partial x_s,
\end{equation}
\begin{equation}\label{eq_A22}
\bm F_s^{\rm GG} \equiv -\frac{1}{2}q^2\partial\bm C_{\rm{G}}^{-1}/\partial x_s.
\end{equation}
\noindent Here, $\bm F_s^0$, $\bm F_s^{\rm G}$ and $\bm F_s^{\rm GG}$ are matrices of constant forces that is related to geometry and material (the matrix dimension is consistent with their multipliers ${\bm n}$ or $\bm n_{\rm G}$ to make $F_s$ a scalar); $\bm \alpha_s$ and $\bm \alpha_s^{\rm G}$ are constant vectors of length reciprocal. These parameters come from the spatial derivative of capacitance matrix.
Note that $\alpha_{ss}$, the $s$\textsuperscript{th} component of the vector $\bm \alpha_s$, is usually much larger than other components. If $\bm n_{\rm G}=0$, we could drop the small terms and use $F_s \cong q \tilde{n}_s \alpha_{ss}V(t)$, which is linear to $\tilde{n}_s$ and consistent with the assumption in \cite{Ahn2006_PRL}. For large $\bm n_{\rm G}$, $F_s \cong q \bm n_{\rm G} \cdot \bm \alpha_s^{\rm G} V(t)$ is a fair approximation.

The white noise can be added to Eq.~(\ref{eq_1}) to account for heating effects, but its energy $k_{\rm B} T/2$ is usually far too small compared to the electric driving vibrations.

The change of energy for an electron tunneling forward/backward through the $j$\textsuperscript{th} junction ($j=1,2,3$) is $\Delta E_j^{\pm} = E_{\rm C}|_{\bm Q_{\rm S},Q_0} - E_{\rm C}|_{\bm Q'_{\rm S},Q'_0}$, where $\bm Q'_{\rm S} = \bm Q_{\rm S} \mp q\bm T_j$, and $Q'_0 = Q_0$ for $j=1,2$ whereas $Q'_0 = Q_0 \mp q$ for $j=3$. Using (\ref{eq_A15}), we obtain
\begin{equation}\label{eq_A23}
\Delta E_j^{\pm} = \mp q(\bm T_j \bm C_{\rm S}^{-1})\cdot\bm Q_{\rm S} + \frac{1}{2}q^2\bm T_j \bm C_{\rm S}^{-1}\bm T_j^{\rm T} \pm q\bm T_j\bm C_{\rm S}^{-1}\bm B^{\rm T}\bm Q_{\rm G}.
\end{equation}

Another question is how much work the external voltage source spends if an electron tunnels from a metal island to a neighbor island. If $\bm Q_{\rm S}$ changes by $\Delta \bm Q_{\rm S}$, according to (\ref{eq_A10}), the change of $Q_0$ is
\begin{equation}\label{eq_A24}
\Delta Q_0 = \Delta \bm Q_{\rm S} \bm \zeta^{\rm T},
\end{equation}

\noindent which is the amount of charge pumped through the external source back to the source electrode. The process can be seen as instantaneous, because $\Delta Q_0$ represents the mean of charge variation for single electron tunneling and it is the same order of infinitesimal of $\Delta t$. The work consumed is $W = \Delta Q_0 V$. Thus, the work done by the external circuit to pump an electron forward through the $j$\textsuperscript{th} junction ($j=1,2,3$) is
\begin{equation}\label{eq_A25}
W_j = qV\kappa_j,
\end{equation}

\noindent where $\kappa_j$ is the $j$\textsuperscript{th} element of the vector
\begin{equation}\label{eq_A26}
\bm{\kappa} = [-\zeta_1, \zeta_1 - \zeta_2, \zeta_2 + 1],
\end{equation}

\noindent and is a unitless coefficient representing the number of electrons pumped between the electrodes by an outside voltage source when an electron tunnels through the $j$\textsuperscript{th} junction, satisfying $\kappa_1 + \kappa_2 + \kappa_3 = 1$. The backward work is just the opposite value of (\ref{eq_A24}). Thus, the change of the free energy for $\Delta t \rightarrow 0$ is
\begin{eqnarray}\label{eq_A27}
U_j^{\pm} &=& \pm W_j - \Delta E_j^{\pm} \nonumber\\
&=& -\frac{1}{2} q^2 \bm T_j \bm C_{\rm{S}}^{-1} \bm T_j^{\rm T}
\pm [q(\bm T_j \bm C_{\rm{S}}^{-1}) \cdot (\bm Q_{\rm{S}} - \bm Q_{\rm{G}}\bm B) + qV\kappa_j].
\end{eqnarray}

\noindent Defining $\bm E^0 = [E_1^0, E_2^0, E_3^0]$ with $E_j^0 = \frac{1}{2} q^2 \bm T_j \bm C_{\rm{S}}^{-1} \bm T_j^{\rm T}$ to denote the ground-state energies, we have
\begin{equation}\label{eq_8}
U_j^{\pm}(\bm n, t) = -E_j^0 \mp (\bm E^0 \bm \Theta_j)\cdot(\bm n-\bm n_{\rm G}\bm B) \pm q\kappa_jV(t),
\end{equation}

\noindent where $\bm B$ is a unitless matrix defined in (\ref{eq_A7}) to describe the effect of gate bias, and $\bm\Theta_j$ is a constant matrix that links the energy and electron numbers:
\begin{equation}\label{eq_A29}
\bm \Theta_1 \equiv
\left[ \begin{array}{rr}
2 & 1 \\
0 & -1 \\
0 & 1
\end{array} \right]
,\;\;\bm \Theta_2 \equiv
\left[ \begin{array}{rr}
-1 & -1 \\
-1 & 1 \\
1 & 1
\end{array} \right]
,\;\;\bm \Theta_3 \equiv
\left[ \begin{array}{rr}
-1 & 0 \\
1 & 0 \\
-1 & -2
\end{array} \right].
\end{equation}

\noindent This is the unperturbed $U_j^{\pm}$ for mechanical motions.
Note that $E_j^0$, $\bm B$ and $\kappa_j$ can be described by the capacitance matrix which depends on geometry and material.

For the first-order perturbation for $\bm x$, ${E}_{\rm C}$ should be subtracted by $\bm F(\bm n,t)\cdot \bm x$, where $\bm F=[F_1,F_2]$ is the force vector on shuttles. For $\bm n_{\rm G}=0$, we have
\begin{displaymath}\label{eq_02}
U_j^{\pm} (\bm n,\bm x,t) \cong U_j^{\pm}(\bm n,t) \mp [q V(t) \bar{\bm{\alpha}}_j + \bm n \bar{{\bm F}}^0_j] \cdot \bm x, \nonumber
\end{displaymath}
\noindent where $\bar{{\bm F}}^0_j = [\bm{F}^0_1 \bm{T}_j^{\rm T},\bm{F}^0_2 \bm{T}_j^{\rm T}]$, $\bar{\bm{\alpha}}_j = {\bm T}_j [\bm{\alpha}_1^{\rm T},\bm{\alpha}_2^{\rm T}]$. For $\bm n_{\rm G}\neq0$, we further add $ \mp {\bm T}_j [\bm F_1^{\rm G} \bm n_{\rm G}^{\rm T}, \bm F_2^{\rm G} \bm n_{\rm G}^{\rm T}] \cdot \bm x$ to $U_j$.


\section{Approximation of the Master Equation}\label{sec5}

Although the linear master equation (\ref{eq_9}) is meaningful for analysis, it is still difficult to solve numerically with high accuracy, even if $\bm n$, $\bm x$ and $\bm v$ can be well truncated and sampled, since the PDF has 7 variables. A feasible approach is the Monte-Carlo method, which relies on repeated random sampling of the sample space. This method is based on the conditional probability from $t$ to $t' = t + \Delta t$ with $\Delta t$ being a small time step. Ideally $\Delta t \rightarrow 0$, $P(x'_s,v'_s,t'|\bm n,\bm x,\bm v,t)$ is a delta function given by (\ref{eq_2}). Thus, for a specific sample, we can assume that the change of $\tilde{\bm x}(t)$ and $\tilde{\bm v}(t)$ are uniquely determined by the system state at time $t$, whereas the change of $\tilde{\bm n}(t)$ (i.e., $\bm \mu \bm T$) is determined by the random numbers produced according to $P(\mu_j,t'|\bm n,\bm x,\bm v,t)$ given by (\ref{eq_6}), as long as $\Delta t$ is sufficiently small. The accuracy of the Monte-Carlo method is also related to the sampling size. However, as the sample space grows with the time evolution, which is a necessary wait for the PDF to fall into the periodic steady state, a huge number of samples is usually needed for accuracy of the result.

The Monte-Carlo method has several advantages: First, it is straightforward to give credible probability distribution, directly from the physics intuition rather than complex mathematical derivation. Second, it can record not only the time evolution of the mean and variance of random variables but also the covariance of random variables at different time. Third, the load is easy to control via the sampling size, which can be progressively increased for higher accuracy. We actually used the Monte-Carlo method as a start point to investigate the properties of the solution and compare to our other numerical results, e.g., the solution's proximity of Gaussian distribution is shown in Fig. \ref{fig2}. However, the disadvantages include that the computation cost is high and the result contains considerable noise. Because the direct current is the average of current flowing toward different directions, it is usually hidden by the noise in the calculated current. Therefore, although the Monte-Carlo method is good for rough result that reveal the motion features, it is ineffective for device simulation.

Another idea is to reduce the number of variables in (\ref{eq_9}) and adopt the Method of Lines that replaces the spatial derivatives by central finite differences in order to convert (\ref{eq_9}) into time-domain ordinary differential equations, each for a spatial sampling point. Thus, we can ignore the generally weak correlation of $\tilde{\bm n}(t)$ and $\tilde{\bm x}(t)$ and the relatively small variance of $\tilde{\bm x}(t)$. Then, by integrating (\ref{eq_9}) over $x_1,x_2,v_1,v_2$, we can transform it into an equation for the marginal distribution $P(\bm n,t)=\int\! P(\bm n,\bm x,\bm v,t){\rm d}\Omega$:
\begin{eqnarray}\label{eq_11}
&&\frac{\partial P(\bm n,t)}{\partial t}= \sum_{j = 1}^{3}\sum_{\pm}\left(\hat{\mathcal{N}}_{\mp\bm T_j} - 1\right) \Gamma_j^{\pm}(\bm n,\langle\tilde{\bm x}\rangle,t) P(\bm n,t),
\end{eqnarray}

\noindent where $\langle\tilde{\bm x}\rangle$ is linked to $\langle F_s(\tilde{\bm n},t)\rangle$ by the mean of (\ref{eq_1}). In fact, this is equivalent to the nonlinear master equation formulated in in pervious studies\cite{Ahn2006_PRL, Prada2012_PRB}. The nonlinearity is due to the implicit relation of $\langle\tilde{\bm x}\rangle$ and $P(\bm n,t)$. Without linearity, we can hardly discuss the solution analytically, since superposition of initial conditions
and solutions are prohibited. For numerical solution, there are also many disadvantages of adopting this equation: PDF also is not well normalized; convergence and accuracy with a long time evolution is a challenge; the computation load is still costly if the range of $\bm n$ in consideration is large; the approximation of small variance and covariance does not hold in the resonance scenarios where ${\rm max}{\langle\tilde{\bm x}(t)\rangle} \gg \lambda_j$.

\section{Equations for Means and Variances}

We are most interested in knowing the measurable physical quantities which are actually the mean values of the random variables, rather than the PDF. Actually, without solving the PDF, we can build deterministic equations for the means from (\ref{eq_9}) and solve the means as time-dependent functions. Multiply both sides of (\ref{eq_9}) by $n_s$ ($s=1,2$), and sum over $\bm n$, $\bm x$, $\bm v$. We can then exchange the sequence of summation, integration and differential, and obtain
\begin{eqnarray}\label{eq_E1}
\frac{\partial}{\partial t}\sum_{\bm n}n_s \int\! P{\rm d}\Omega
&&= \sum_{j=1}^3 \sum_{\bm n} \int\!{\rm d}\Omega \sum_\pm
\left[n_s \hat{\mathcal{N}}_{\mp\bm T_j}(\Gamma_j^{\pm} P) - n_s \Gamma_j^{\pm} P\right] \nonumber\\
&&- \sum_{\bm n}n_s\sum_{s'=1}^2\int\!{\rm d}\Omega
\left[-  \frac{\partial (v_{s'} P)}{\partial x_{s'}}
+\frac{\partial}{\partial v_{s'}} \left(\gamma_{s'} v_{s'}+\omega_{s'}^2 x_{s'}-\frac{F_{s'}(\bm n,t)}{m_{s'}} \right)P\right].
\end{eqnarray}
\noindent The left side is namely the time derivative of $\langle \tilde{n}_s \rangle$. The second term of the right side is zero, because the integral of partial derivative is determined by the boundary condition which is obviously zero. For the first term of the right side, we have $n_s \hat{\mathcal{N}}_{\mp\bm T_j}(\Gamma_j^{\pm} P) = \hat{\mathcal{N}}_{\mp\bm T_j} (n_s \Gamma_j^{\pm} P) \pm (T_{js} \Gamma_j^{\pm} P)$. Since the operator $\hat{\mathcal{N}}_{\mp\bm T_j}$ makes no effect inside the summation over $\bm n$, the first item gets neutralized by $- n_s \Gamma_j^{\pm} P$, and only the summation of $\pm T_{js} \Gamma_j^{\pm} P$ remains in this term, i.e., $T_{js} (+\Gamma_j^{+}-\Gamma_j^{-}) P$ before summation. By defining $\Gamma_j \equiv \Gamma_j^+  - \Gamma_j^- $, we hence obtain
\begin{equation}\label{eq_19}
\frac{{\rm d} \langle \tilde{n}_s \rangle}{{\rm d} t} = \sum_{j=1}^{3}{T_{js} \langle \Gamma_j(\tilde{\bm n},\tilde{\bm x},t)\rangle}.
\end{equation}

Following similar procedures, we can calculate the time derivative of $\langle x_s \rangle$ and $\langle v_s \rangle$: Multiply both sides of (\ref{eq_9}) by $x_s$ ($s=1,2$), and sum over $\bm n$, $\bm x$, $\bm v$; we have
\begin{eqnarray}\label{eq_E2}
\frac{{\rm d} \langle \tilde{x}_s \rangle}{{\rm d} t}
&&= \sum_{j=1}^3 \sum_{\bm n} \sum_\pm\int\!{\rm d}\Omega
x_s (\hat{\mathcal{N}}_{\mp\bm T_j} - 1) \Gamma_j^{\pm} P \nonumber\\
&&- \sum_{\bm n} \sum_{s'=1}^2\int\!{\rm d}\Omega
\left[- x_s \frac{\partial (v_{s'} P)}{\partial x_{s'}}
+x_s \frac{\partial}{\partial v_{s'}} \left(\gamma_{s'} v_{s'}+\omega_{s'}^2 x_{s'}-\frac{F_{s'}(\bm n,t)}{m_{s'}} \right)P\right].
\end{eqnarray}
\noindent Obviously, the first term of the right side is zero since the operator $\hat{\mathcal{N}}_{\mp\bm T_j}$ can be canceled with the summation. In the second term, the only non-zero integrand is $x_s {\partial (v_{s'} P)}/{\partial x_{s'}}$ for $s' = s$ using the following relation
\begin{equation}\label{eq_E3}
x_s \frac{\partial (v_{s} P)}{\partial x_{s}} = \frac{\partial (x_s v_{s} P)}{\partial x_{s}} - v_s P.  \nonumber\\
\end{equation}

\noindent Thus, the total nonzero integrand after summing over $s'$ is $v_s P$, making the total right side as $\langle v_s \rangle$. We hence obtain
\begin{equation}\label{eq_E4}
\frac{{\rm d} \langle \tilde{x}_s \rangle}{{\rm d} t} = \langle \tilde{v}_s \rangle.
\end{equation}

Multiply both sides of (\ref{eq_9}) by $v_s$ ($s=1,2$), and sum over $\bm n$, $\bm x$, $\bm v$; we have
\begin{eqnarray}\label{eq_E5}
\frac{{\rm d} \langle \tilde{x}_s \rangle}{{\rm d} t}
&&= \sum_{j=1}^3 \sum_{\bm n} \sum_\pm\int\!{\rm d}\Omega
v_s (\hat{\mathcal{N}}_{\mp\bm T_j} - 1) \Gamma_j^{\pm} P \nonumber\\
&&- \sum_{\bm n} \sum_{s'=1}^2\int\!{\rm d}\Omega
\left[- v_s \frac{\partial (v_{s'} P)}{\partial x_{s'}}
+v_s \frac{\partial}{\partial v_{s'}} \left(\gamma_{s'} v_{s'}+\omega_{s'}^2 x_{s'}-\frac{F_{s'}(\bm n,t)}{m_{s'}} \right)P\right].
\end{eqnarray}
\noindent Similarly, the first term of the right side is zero. In the second term, the non-zero integrand is the second part with $s' = s$, and after the summation of $s'$ this term becomes
\begin{equation}\label{eq_E6}
v_s \frac{\partial}{\partial v_{s}} \left(\gamma_{s} v_{s}+\omega_{s}^2 x_{s}-\frac{F_{s}(\bm n,t)}{m_{s}} \right)P
= \frac{\partial}{\partial v_{s}} v_s \left(\gamma_{s} v_{s}+\omega_{s}^2 x_{s}-\frac{F_{s}(\bm n,t)}{m_{s}} \right)P
- v_s \left(\gamma_{s} v_{s}+\omega_{s}^2 x_{s}-\frac{F_{s}(\bm n,t)}{m_{s}} \right)P.
\end{equation}
Since the first part should be zero after integration, (\ref{eq_E5}) becomes
\begin{equation}\label{eq_E7}
\frac{{\rm d} \langle \tilde{x}_s \rangle}{{\rm d} t}
= \gamma_{s} {\langle \tilde{v}_s \rangle} + \omega_{s}^2 {\langle \tilde{x}_s \rangle} - \frac{\langle F_{s}(\tilde{\bm n},t)\rangle}{m_{s}}.
\end{equation}
\noindent It is conceivable that (\ref{eq_E4}) and (\ref{eq_E7}) are the mean of (\ref{eq_1}), and they can also be written in a vector form.

We want to make use of (\ref{eq_19}), (\ref{eq_E4}) and (\ref{eq_E7}) to calculate the means of variables, instead of solving the PDF from (\ref{eq_9}). However, the difficulty is to describe $\langle \Gamma_j(\tilde{\bm n},\tilde{\bm x},t)\rangle$ in terms of the mean and variance of the random variables. Fortunately, the correlation between $\tilde{\bm n}(t)$ and $\tilde{\bm x}(t)$ is usually rather weak (the covariance is much smaller than the geometric mean of variances of the two variables).
For every infinitesimal time step, since the electron tunneling and mechanical motion are independent events and the conditional probability is the product of the probability of the two events, the infinitesimal change of $\tilde{\bm x}(t)$ is independent of the change of $\tilde{\bm n}(t)$, but it is correlated to the state variable $\tilde{\bm n}(t)$ which is further correlated to the state variable at a previous time. After evolution of certain time, $\tilde{\bm n}(t)$ and $\tilde{\bm x}(t)$ are both correlated to the process ${\tilde{\bm n}(t')|t'<t}$. From the solution of the stochastic differential equation given in (\ref{eq_0a}) where $\tilde{\bm x}(t)$ is described by $\tilde{\bm n}(t)$, we know the correlation is decided by the correlation of $\tilde{\bm n}(t)$ and $\tilde{\bm n}(t'), t'<t$. It is small as long as the electron jump at different time are weakly dependent. This can be verified by the maximum of $\rm{Cov}[n_s(t),n_{s'} (t')]$ in the periodic steady state for $|t-t'|\geq \pi/\omega$, using the model with the correlation of $\tilde{\bm n}$ and $\tilde{\bm x}$ ignored. In addition, the variance of $\tilde{x_s}(t)$ is also decided by this correlation. It can also be ignored if the maximum is sufficiently small (say 0.05). These approximations, if valid, can greatly simplify the model and reduce the computation load. Thus, from simple to complex, we can develop three following models.

\subsection{Circuit Model}\label{sec6}

First, let us consider a simple approximation: Use the unperturbed $U_j^\pm (\bm n,t)$ to replace $U_j^\pm (\bm n,\bm x,t)$ and assume
\begin{equation}\label{eq_02}
\Gamma_j (\tilde{\bm n},t) \cong U_j (\tilde{\bm n},t)/(q^2 R_j^0),
\end{equation}

\noindent where
\begin{eqnarray}\label{eq_03}
U_j (\tilde{\bm n},t) &=& [U_j^+ (\tilde{\bm n},t) - U_j^- (\tilde{\bm n},t)]/2 \nonumber\\
&=& q\kappa_j V(t) - (\tilde{\bm n}-\bm n_{\rm G}\bm B) \cdot (\bm E^0 \bm \Theta_j).
\end{eqnarray}

\noindent This is valid when $|U_j| \gg k_{\rm{B}} T$. Under this assumption, $U_j$ stands for the voltage drop through the $j$\textsuperscript{th} junction and is proportional with the tunneling current $\langle q \Gamma_j\rangle$. This junction voltage $U_j$ is is a linear function of total applied voltage $V(t)$ and the charge number $\tilde{\bm n}$, and so is the the current.

In this approximation, we also ignore the correlation of $\tilde{\bm n}(t)$ and $\tilde{\bm x}(t)$, so that $\langle \Gamma_j(\bm n, \bm x, t) \rangle= \langle K_j(\bm x)\rangle \langle \Gamma_j(\bm n,t)\rangle$. This is usually valid for non-resonant cases. Additionally, we ignore the variance of $\tilde{\bm x}(t)$ and assume $\langle K_j(\bm x)\rangle =K_j (\langle \bm x\rangle)$. These two assumptions are correlated, because randomness of the system comes from electron tunneling represented by $\tilde{\bm n}(t)$. The stand deviation of $\tilde{\bm x}(t)$ is usually much smaller than its mean and can be ignored in calculation.

Then, we can replace (\ref{eq_19}) by
\begin{equation}\label{eq_20}
\frac{{\rm d} \langle \tilde{n}_s \rangle}{{\rm d} t} \cong \sum_{j=1}^{3}{\frac{T_{js}}{q^2 R_j^0} {\rm e}^{-\langle\tilde{\bm x} \rangle\cdot \bm T_j/\lambda_j} U_j(\langle\tilde{\bm n}\rangle,t)}.
\end{equation}

\noindent This equation can be combined with (\ref{eq_E4}) and (\ref{eq_E7}) to solve $\langle\tilde{\bm x}\rangle$ and $\langle\Gamma_j\rangle$. In fact, (\ref{eq_20}) can be interpreted as a circuit model, because the variation of charges ${{\rm d} \langle \tilde{n}_s \rangle}/{{\rm d} t}$ being proportional to the junction voltage $U_j$ represents the capacitance relation. If $\partial\langle\tilde{n}_s\rangle/\partial t \approx 0$ which probably happens at low frequencies, (\ref{eq_19}) falls to a simpler circuit model for the adiabatic limit proposed before~\cite{Ahn2006_PRL}, which ignores the charging current of capacitors, but it does not hold for high frequency excitation. The circuit model does not take into account the variance of variables and it is also too rough for simulation of the current.

\subsection{Variance of $\tilde{\bm n}$ and $\tilde{\bm x}$ Considered}\label{sec6}

For higher accuracy, we assume $\tilde{n}_s$ ($s = 1, 2$) have the bivariate normal distribution. The covariance matrix of $\tilde{\bm n}$ is denoted by $\bm D$, i.e., $D_{ss}(t) = \langle \tilde{n}_s^2 \rangle - \langle \tilde{n}_s \rangle^2$ is the variance of $\tilde{n}_s$, and $D_{12}(t) = \langle \tilde{n}_1 \tilde{n}_2 \rangle - \langle \tilde{n}_1 \rangle \langle \tilde{n}_2 \rangle$ is the covariance of $\tilde{n}_1$ and $\tilde{n}_2$. Nevertheless, we still ignore the correlation of $\tilde{\bm n}$ and $\tilde{\bm x}$ and the variance of $\tilde{\bm x}$ as in the circuit model. The Taylor expansion of $\Gamma_j^{\pm}(\tilde{\bm n}, \tilde{\bm x},t)$ for $\tilde{\bm n}$ and $\tilde{\bm x}$ around $\langle\tilde{\bm n}\rangle$ and $\langle\tilde{\bm x}\rangle$ gives
\begin{eqnarray}\label{eq_21}
\langle\Gamma_j^{\pm}(\tilde{\bm n}, \tilde{\bm x},t)\rangle \cong \frac{1}{q^2 R_j^0} \langle K_j(\tilde{\bm x})\rangle
\sum_{l_1,l_2 = 0}^{\infty}\frac{(\mp 1)^{l_1+l_2}}{l_1!\,l_2!} E_{1j}^{l_1} E_{2j}^{l_2} \cdot Y_{l_1+l_2}\left(U_j^\pm (\langle {\bm n}\rangle, \langle {\bm x}\rangle, t)\right)  M_{l_1 l_2},
\end{eqnarray}

\noindent where $\langle K_j(\tilde{\bm x})\rangle \cong K_j(\langle\tilde{\bm x}\rangle)$, $Y_{l}(U) = \partial^l \left[U/(1-{\rm e}^{-U/k_{\rm{B}}T}) \right]/\partial U^l$, $E_{1j}$ and $E_{2j}$ are components of $\bm E^0\bm\Theta_j$, and $M_{l_1 l_2}$ is the $l_1,l_2$-order mixed moment of $n_1,n_2$ following the Isserlis' theorem (see Appendix-\ref{App_B}):
\begin{eqnarray}\label{eq_22}
M_{l_1 l_2} \equiv \langle(\tilde{n}_1 - \langle \tilde{n}_1 \rangle)^{l_1}(\tilde{n}_2 - \langle \tilde{n}_2 \rangle)^{l_2}\rangle
= \left\{ \begin{array}{lr}
\displaystyle{\!\sum\limits_{k = 0}^{l_{\rm m}}}{\frac{l_1!\,l_2! D_{11}^{\frac{l_1}{2} - k}D_{22}^{\frac{l_2}{2} - k}}{(l_1 - 2k)!!(l_2 - 2k)!!}\frac{D_{12}^{2k}}{(2k)!}}, & \!\!\!\!l_1,l_2 \rm{\ even}, \\
\displaystyle{\!\sum\limits_{k = 0}^{l_{\rm m}}}{\frac{l_1!\,l_2! D_{11}^{\frac{l_1-1}{2} - k}D_{22}^{\frac{l_2-1}{2} - k}}{(l_1-1-2k)!!(l_2-1-2k)!!}\frac{D_{12}^{2k+1}}{(2k+1)!}}, & \!l_1,l_2 \rm{\ odd}, \\
\;0, & \!\!\!\!\!\!\!l_1+l_2 \rm{\ odd},
\end{array} \right. \nonumber
\end{eqnarray}

\noindent where $l_{\rm{m}}$ denotes the integer part of $\min(l_1, l_2)/2$. Similar to building (\ref{eq_19}) for $\frac{{\rm d} n_s}{{\rm d} t}$, we can also build equations for the variance and covariance (see details in Appendix-\ref{App_A}):
\begin{eqnarray}\label{eq_23}
\frac{{\rm d} D_s}{{\rm d} t} &=& \sum_{j=1}^{3}[2T_{js}\langle \Gamma_j(\tilde{\bm n},\tilde{\bm x},t)(\tilde{n}_s - \langle\tilde{n}_s\rangle)\rangle
+ T_{js}^2 \langle \Gamma_j^*(\tilde{\bm n},\tilde{\bm x},t)\rangle],
\end{eqnarray}

\begin{eqnarray}\label{eq_24}
\frac{{\rm d} D_{12}}{{\rm d} t} &=& \sum_{j=1}^{3}\sum_{s=1}^{2}T_{j(3-s)}\langle \Gamma_j(\tilde{\bm n},\tilde{\bm x},t)(\tilde{n}_s - \langle\tilde{n}_s\rangle)\rangle
- \langle \Gamma_2^*(\tilde{\bm n},\tilde{\bm x},t)\rangle,
\end{eqnarray}
\noindent where $\Gamma_j^*(\tilde{\bm n},t) = \Gamma_j^+ (\tilde{\bm n},t) + \Gamma_j^- (\tilde{\bm n},t)$, and the result $\langle\Gamma_j(\tilde{\bm n},t)(\tilde{n}_s - \langle\tilde{n}_s\rangle)\rangle$ is similar to the right side of Eq.~(\ref{eq_21}) with $M_{l_1 l_2}$ replaced by $M_{l_1+1, l_2}$ or $M_{l_1, l_2+1}$ according to $s=1,2$.

\subsection{Weak Correlation of $\tilde{\bm n}$ and $\tilde{\bm x}$}\label{sec6}
Now consider the weak correlation of $\tilde{\bm n}$ and $\tilde{\bm x}$.
Assume $\tilde{x}_s$, $\tilde{v}_s$ and $\tilde{n}_s$ ($s = 1, 2$) have the multivariate normal distribution. $\Lambda ¦«_{ss}$ denotes the variance of $\tilde{x}_s$; $W_s$ denotes the variance of $\tilde{v}_s$; $\Sigma_s$ denotes the covariance of $\tilde{x}_s$ and $\tilde{v}_s$; $X_s$ denotes the covariance of $\tilde{x}_s$ and $\tilde{n}_s$; $Y_s$ denotes the covariance of $\tilde{n}_s$ and $\tilde{v}_s$.
Other covariances which are between different shuttles are generally negligible, due to small correlation between $F_s(\tilde{\bm n},t)$ and $\tilde{n}_{s'}$ ($s \neq s'$), i.e., $\alpha_{ss} \gg \alpha_{ss'}$.
We assume these covariances are small and only consider the first order approximation, which means the products of covariances are ignored in the following derivation.

Multiply (\ref{eq_9}) with $x_s^2$ and sum over $\bm n$, $\bm x$, $\bm v$, thus we can derive ${\rm d}\langle \tilde{x}_s^2\rangle/{\rm d}t=2\langle\tilde{x}_s \tilde{v}_s\rangle$. Combine it with (\ref{eq_E4}) and consider $\Lambda_{ss}=\langle \tilde{x}_s^2\rangle - {\langle \tilde{x}_s\rangle}^2$; we get
\begin{equation}\label{eq_13}
\frac{{\rm d} \Lambda_{ss}}{{\rm d} t} = 2\Sigma_s
\end{equation}

\noindent Similarly, we can build equations describing the time derivative of $W_s$, $\Sigma_s$, $X_s$, and $Y_s$ (see details in Appendix-\ref{App_A} and \ref{App_C}):
\begin{equation}\label{eq_14}
\frac{{\rm d}W_{s}}{{\rm d} t} = 2\left[-\gamma_s W_s-\omega_s^2 \Sigma_s + Y_s\frac{\langle f_s(\tilde{\bm n},t)\rangle}{m_s}\right],
\end{equation}

\begin{equation}\label{eq_15}
\frac{{\rm d}\Sigma_{s}}{{\rm d} t} = W_s-\gamma_s \Sigma_s-\omega_s^2 \Lambda_{ss} + X_s\frac{\langle f_s(\tilde{\bm n},t)\rangle}{m_s},
\end{equation}

\begin{equation}\label{eq_16}
\frac{{\rm d} X_{s}}{{\rm d} t} = \sum_{j=1}^{3}T_{js}\langle K_j(\tilde{\bm x})\rangle
\left[g_{js}(t)X_s-\frac{T_{js}}{\lambda_j}G_j (t)\Lambda_{ss}\right] + Y_s,
\end{equation}

\begin{eqnarray}\label{eq_17}
\frac{{\rm d} V_{ss}}{{\rm d} t} =
\sum_{j=1}^{3}T_{js}\langle K_j(\tilde{\bm x})\rangle
\left[g_{js}(t)Y_s-\frac{T_{js}}{\lambda_j}G_j (t)\Sigma_s\right]
- \gamma_s Y_s - \omega_s^2 X_s + \frac{\langle F_s(\tilde{\bm n},t)(\tilde{n}_s-\langle \tilde{n}_s \rangle)\rangle}{m_s}.
\end{eqnarray}

\noindent where $\langle K_j(\tilde{\bm x})\rangle = K_j(\langle\tilde{\bm x}\rangle){\rm e}^{{\bm T}_j\bm \Lambda{\bm T}_j^{\rm T} \lambda_j^2}$, and we define $f_s (\bm n,t) \equiv {\partial F_s (\bm n,t)}/{\partial n_s}$ which is a linear function of $\bm n$,
$\bm g_j(t) = \langle\partial\Gamma_j(\tilde{\bm n},t)/\partial \tilde{\bm n}\rangle$ with $g_{js}(t)$ as the $s$\textsuperscript{th} component, and
\begin{eqnarray}\label{eq_18}
G_j(t)=\langle\Gamma_j(\tilde{\bm n},t)\rangle -\bm g_j(t){\rm diag}(\bm T_j)[X_1,X_2]^{\rm T}/\lambda_j,
\end{eqnarray}
wherein ${\rm diag}(\bm T_j)$ is the diagonal matrix with $\bm T_j$ as the diagonal.

Thus, we can combine the ordinary differential equations (\ref{eq_19}),(\ref{eq_21})-(\ref{eq_17}) and the mean of (\ref{eq_1}) in order to solve the mean of variables. The equations can be solved numerically by the Euler's method or the improved Euler's method. Empirically, the order of moment in (\ref{eq_22}) can be set to $l_1+l_2 \leq 4$ for decent accuracy.

\section{Nonlinear Properties}\label{sec7}

To perfect the model, we can consider nonlinear resistance and nonlinear vibration. According to the calculation for tunnel-junction~\cite{Simmons1963_1_JAP}, the conductance $1/R_j^0$ in (\ref{eq_7}) could additionally have a factor of $(1+\beta_j U_j^2)$, where $\beta\propto(d_j/\lambda_j)^2$ if the length of tunnel junction $d_j \gg \lambda_j$,  In addition, we can add the terms $k_2x_s^2$ and $k_3x_s^3$ in (\ref{eq_1}) to describe the nonlinearity of nanopillar vibration.


\section{Direct current}\label{sec8}

The macroscopic current can be calculated from the mean rate of electrons tunneling:
\begin{equation}\label{eq_25}
I(t) = C_0 \frac{{\rm d} V}{{\rm d} t} + q\sum_{j=1}^{3}{\kappa_j \langle \Gamma_{j}(\tilde{\bm n},\tilde{\bm x}, t)\rangle},
\end{equation}

\noindent where the positive current indicates flowing from right to left in Fig.\ref{fig1_a},and $C_0$ is the equivalent capacitance seen from the electrode. In the periodic steady-state solution, $I(t)$ is periodic, so the direct current is a time-average of $I(t)$ for a full period. If $\langle \tilde{n}_1 \rangle$, $\langle \tilde{n}_2 \rangle$ and $V(t)$ are periodic, we can substitute (\ref{eq_19}) into (\ref{eq_25}) and remove the terms with $\partial\langle \tilde{n}_{1,2}\rangle/\partial t$. Thus,
\begin{equation}\label{eq_27}
I_{\rm{dc}} = \frac{q\omega}{2\pi} \int_{t_0}^{t_0+2\pi/\omega}\!{\langle \Gamma_j (\tilde{\bm n}, \tilde{\bm x}, t)\rangle {\rm d} t}
\end{equation}
\noindent where $j=1$,2 or 3 makes no difference. The value of $I_{\rm{dc}}$ is usually small because the current flows in another direction after half a period $\pi/\omega$.

An important conclusion about symmetry breaking of current can be derived from (\ref{eq_9}). If $\bm n_{\rm G}=0$ and $V(t) = -V(t + \pi/\omega)$, from (\ref{eq_7}) and (\ref{eq_8}) we obtain
\begin{equation}\label{eq_28a}
\Gamma_j^\pm(\bm n,\bm x,t + \pi/\omega) = \Gamma_j^\mp(-\bm n,\bm x,t).
\end{equation}

\noindent and from (\ref{eq_A17}) we obtain
\begin{equation}\label{eq_28b}
F_s (\bm n,t + \pi/\omega) = F_s (-\bm n,t)
\end{equation}
Let us replace $t$ by $t + \pi/\omega$ in (\ref{eq_9}), and substitute the above relations into (\ref{eq_9}) with $\bm n$ replaced by $-\bm n$. Then, we obtain the same equation for $P(-\bm n,\bm x,\bm v,t + \pi/\omega)$:
\begin{eqnarray}\label{eq_29}
&&\frac{\partial P(-\bm n,\bm x,\bm v,t + \frac{\pi}{\omega})}{\partial t}
= \sum_{j=1}^{3}\sum_{\pm}[\Gamma_j^\mp(\bm n\pm\bm T_j,\bm x,t)
P(-\bm n\mp\bm T_j,\bm x,\bm v,t + \frac{\pi}{\omega})
- \Gamma_j^\mp (\bm n,\bm x,t) P(-\bm n,\bm x,\bm v,t + \frac{\pi}{\omega})] \nonumber\\
&&+\sum_{s = 1}^{2} \left[ \gamma_s P(-\bm n,\bm x,\bm v,t)
- v_s \frac{\partial P(-\bm n,\bm x,\bm v,t + \frac{\pi}{\omega})}{\partial x_s}
+ \left(\gamma_s v_s+\omega_s^2 x_s-\frac{F_s(\bm n,t)}{m_s}\right)
\frac{\partial P(-\bm n,\bm x,\bm v,t + \frac{\pi}{\omega})}{\partial v_s} \right]
\end{eqnarray}

\noindent Changing the sequence of `$+$' and `-' for the summation , this is actually the same equation as (\ref{eq_B11}) except that the unknown function is changed from $P(\bm n,\bm x,\bm v,t)$ to $P(-\bm n,\bm x,\bm v,t + \pi/\omega)$. Assume such equation has only one periodic steady-state solution which is irrelevant the the initial condition (always normalized). Thus, assuming the equation has a unique periodic steady-state solution, we have
\begin{equation}\label{eq_30}
P(-\bm n,\bm x,\bm v,t + \pi/\omega) = P(\bm n,\bm x,\bm v,t)
\end{equation}

\noindent In this situation, from (\ref{eq_25}) we can derive
\begin{equation}\label{eq_30b}
I(t) = -I(t + \pi/\omega)
\end{equation}
which suggests that no direct current exists without charge or voltage asymmetry, since the current is inverted after exactly half a period.

Thus, there are two ways to break the symmetry:
First is to apply bias on the gate. If $\bm n_{\rm G} \neq 0$, we still have $\Gamma_j^\pm(\bm n,t+\pi/\omega) = \Gamma_j^\mp(2\bm n_{\rm G}\bm B-\bm n,t)$, but $F_s(\bm n,t+\pi/\omega)\neq F_s(2\bm n_{\rm G}\bm B-\bm n,t)$. Thus, $I(t)\neq-I(t+\pi/\omega)$ which enables the direct current.
Second is to introduce even-order harmonics in $V(t)$, which breaks the symmetry of AC voltage after half a period making $V(t) \neq -V(t + \pi/\omega)$. The wave superposition makes use of the nonlinear transport relation. In practice, this could be realized by natural wave distortion, introducing nonlinear elements in the circuit, or magnifying the second-order wave.
Both ways are electric methods. This conclusion is a useful tip not only for the design of electron shuttles but also for general symmetry breaking.

\section{Analytical Estimate}\label{sec9}
We can make a useful analytical estimate in case of $V(t)=V_0\sin(\omega t)$ and small vibrations compared to $\lambda$. We adopt (\ref{eq_20}) with $\langle\bm x\rangle=0$, and assume
\begin{equation}\label{eq_I0}
\langle \tilde{n}_s(t)\rangle = \bm n_{\rm G} \bm B_s + \bar{n}_s \sin(\omega t + \phi),
\end{equation}

\noindent where $\bm B_s$ is the $s$\textsuperscript{th} column of matrix $\bm B$. Thus, we get
\begin{equation}\label{eq_I1}
\frac{{\rm d}\langle\tilde{n}_s(t)\rangle}{{\rm d}t}
\cong \sum_{j=1}^3\frac{T_{js}}{q^2 R_j^0} K_j(\langle\tilde{\bm x}\rangle)
\left[q\kappa_j V(t)-(\bm E^0\bm\Theta_j)\cdot(\langle\tilde{\bm n}(t)\rangle-\bm n_{\rm G}\bm B)\right].
\end{equation}

For small vibrations $\langle\tilde{x}_s\rangle\ll\lambda$, $K_j(\langle\tilde{\bm x}\rangle)\cong1$. With $V(t)=V_0\sin(\omega t)$, $\langle\tilde{\bm n}(t)\rangle$ becomes simple harmonic $\langle\tilde{n}_s(t)\rangle=\bm n_{\rm G}\bm B_s+n_s\sin(\omega t-j\phi)$. Using the phasor analysis, we obtain
\begin{equation}\label{eq_I2}
j\omega \bar{n}_s{\rm e}^{j\omega t}
\cong \sum_{j=1}^3\frac{T_{js}}{q^2 R_j^0}
\left[q\kappa_j V_0-(\bm E^0\bm\Theta_{j1})\bar{n}_1-(\bm E^0\bm\Theta_{j2})\bar{n}_2\right]
{\rm e}^{j\omega t}.
\end{equation}

\noindent For symmetric junctions, $R_1^0=R_3^0$, $E_1^0=E_3^0$, $\kappa_1=\kappa_3$. Consider the value of $\bm\Theta_j$ given in (\ref{eq_A29}) and combine the terms with $\bar{n}_1$ and $\bar{n}_2$. We obtain
\begin{equation}\label{eq_I5}
j\omega \bar{n}_1 q^2
+ \left(\frac{2E_1^0}{R_1^0}+\frac{E_2^0}{R_2^0}\right)\bar{n}_1
+\left(\frac{2E_1^0-E_2^0}{R_1^0}-\frac{E_2^0}{R_2^0}\right)\bar{n}_2
= qV_0 \left(\frac{\kappa_1}{R_1^0}-\frac{\kappa_2}{R_2^0}\right),
\end{equation}
\begin{equation}\label{eq_I6}
j\omega \bar{n}_2 q^2
+ \left(\frac{2E_1^0-E_2^0}{R_1^0}-\frac{E_2^0}{R_2^0}\right)\bar{n}_1
+\left(\frac{2E_1^0}{R_1^0}+\frac{E_2^0}{R_2^0}\right)\bar{n}_2
= qV_0 \left(\frac{\kappa_2}{R_2^0}-\frac{\kappa_1}{R_1^0} \right).
\end{equation}

\noindent By adding the two equations, we obtain $n_1+n_2=0$. Using this relation and $\kappa_2=1-2\kappa_1$, the first equation becomes
\begin{equation}\label{eq_31}
\overline{n}_1 = -\overline{n}_2 = \frac{qV_0}{E_2^0}\frac{\kappa_1-R_1^0/R_{\rm t}^0}{\sqrt{(\omega/\omega_{\rm c})^2 +1}}{\rm e}^{j\phi},
\end{equation}

\noindent where $R_{\rm t}^0 = 2R_1+R_2$ denotes the total resistance, and
\begin{equation}\label{eq_32}
\omega_{\rm c} = \frac{E_2^0}{q^2}\frac{R_{\rm t}^0}{R_1^0 R_2^0}
\end{equation}

\noindent denotes a critical frequency. The angle in (\ref{eq_I0}) is $\phi = \arctan(\omega/\omega_{\rm c})$. Thus, for the low frequency $\omega \ll \omega_{\rm c}$, we have $\overline{n}_1\propto qV_0/E_2^0$ and $\phi\cong 0$, so $\langle\tilde{\bm n}(t)\rangle$ is in phase with $V(t)$ and its magnitude is almost independent of the frequency. However, for the high frequency $\omega \gg \omega_{\rm c}$, we have $\overline{n}_1\propto V_0/\omega$ and $\phi \cong \pi/2$, so the maximum electron numbers decrease with the increase of frequency. Typically, with $E_2^0\sim q^2/C_0\sim 0.01$ eV, $R_1,R_2\sim 1$ G${\rm \Omega}$, the critical frequency is hence in the order of 50 MHz. Thus, the magnitude of $\langle\tilde{\bm n}(t)\rangle$ is rather small for frequencies over 1 GHz, and we need the electric symmetry breaking methods for excitation of effective vibrations. Besides, for device optimization, we should maximize vibrations and the electron numbers. The equation (\ref{eq_31}) also gives us a hint that the geometry can be optimized by maximizing $\kappa_1$ and minimizing $E_2^0$ and $R_1^0/R_2^0$.

For appreciable vibrations, ${\rm e}^{-\langle\bm x\rangle \cdot \bm T_j/\lambda_j}$ is periodic with $\omega$. Thus, $\langle\tilde{\bm n}(t)\rangle$ may have considerable high-order harmonic components $2\omega$, $3\omega$, $\cdots$. This indicates that the resonant frequencies are around $\omega_0/l$ where $l=1,2,3,\cdots$, and the smaller $l$ lead to more substantial vibrations. Note that a small shift of the damped frequency is possible, which depends on many parameters. This phenomenon is known as Arnold's tongues in measurement~\cite{Kim2010_PRL, Kim2012_ACSNano}.

For more appreciable vibrations, $\tilde{\bm x}(t)$ is periodic with frequency $\omega$. If $\omega \approx \omega_1  \approx \omega_2$, $\tilde{\bm x}(t)$ vibrate almost as a simple harmonic wave with frequency $\omega$, since the nanopillars with high quality factors play the role of a resonator. The symmetry breaking can be seen from (\ref{eq_21}) of zeroth order, i.e.,
\begin{equation}\label{eq_I10}
\langle\Gamma_j (\tilde{\bm n}, \tilde{\bm x}, t)\rangle
\cong \frac{1}{q^2 R_j^0} \langle{\rm e}^{-\tilde{\bm x} \cdot \bm T_j/\lambda_j}\rangle
\left[q\kappa_j V(t) - \langle\tilde{\bm n}\rangle \cdot (\bm E^0 \bm \Theta_j)
- \left(q V(t) \bar{\bm{\alpha}}_j + \langle\tilde{\bm n}\rangle \bar{{\bm F}}^0_j \right) \cdot \langle\bm x\rangle \right],
\end{equation}

\noindent where the product of $\langle\tilde{\bm x}\rangle$ and $V(t)$ or $\langle\tilde{\bm n}\rangle$ introduces the DC component and the even order harmonics, which are not inverted after half a period like when $\tilde{\bm n}=0$ is assumed. This means the direct current is related to the magnitude of vibration.

\section{Conclusion}\label{sec4}
We have implemented a full theoretical study on coupled nanomechanical electron shuttles, focusing on the Markovian behavior and the DC output current. By treating the electronic and mechanical motions as stochastic processes, we built a full stochastic model represented by the linear master equation, which enables analysis of symmetry breaking. From this, even-order harmonics of the driving voltage or a gate bias are necessary for observing a DC signal. The simpler nonlinear master equation and circuit model, which were discussed by previous studies, accords with this theory. Beyond, we were able to build the deterministic ordinary differential equations for the mean and covariance of random variables, by assuming the multivariate Gaussian distribution. This provides an efficient method for device-level simulation.

\bibliographystyle{apsrev4-1}
\bibliography{nems_refs}


\begin{appendix}
\section{Mean and Covariances of Variables} \label{App_A}
\begin{eqnarray}\label{eq_B1}
\frac{{\rm d}\langle\tilde{n}_s^2\rangle}{{\rm d}t}
&=& \sum_{j=1}^3\sum_\pm\int\!{\rm d}\Omega
\left[\sum_{\bm n}n_s^2\Gamma_j^\pm(\bm n\mp\bm T_j,t) P(\bm n\mp \bm T_j,\bm x,\bm v,t)
-\sum_{\bm n}n_s^2\Gamma_j^\pm(\bm n,t)P(\bm n,\bm x,\bm v,t)\right]
{\rm e}^{-\bm x\cdot\bm T_j/\lambda_j} \nonumber\\
&=& \sum_{j=1}^3\sum_\pm\int\!{\rm d}\Omega
\sum_{\bm n} \left[(n_s\mp T_{js})^2\pm 2T_{js}(n_s\mp T_{js})+ T_{js}^2\right]
\Gamma_j^\pm(\bm n\mp \bm T_j,t)
P(\bm n\mp\bm T_j,\bm x,\bm v,t)
{\rm e}^{-\bm x\cdot\bm T_j/\lambda_j} \nonumber\\
&&\quad -\sum_{j=1}^3\sum_\pm\int\!{\rm d}\Omega
\sum_{\bm n}n_s^2\Gamma_j^\pm(\bm n,t)
P(\bm n,\bm x,\bm v,t)
{\rm e}^{-\bm x\cdot\bm T_j/\lambda_j} \nonumber\\
&=& \sum_{j=1}^3\sum_\pm\int\!{\rm d}\Omega
\sum_{\bm n} \left[\pm 2T_{js}(n_s\mp T_{js})+ T_{js}^2\right]
\Gamma_j^\pm(\bm n\mp\bm T_j,t)
P(\bm n\mp\bm T_j,\bm x,\bm v,t)
{\rm e}^{-\bm x\cdot\bm T_j/\lambda_j} \nonumber\\
&=& \sum_{j=1}^3\int\!{\rm d}\Omega
\sum_{\bm n}\sum_\pm(\pm 2T_{js}n_s+T_{js}^2)
\Gamma_j^\pm(\bm n,t)
P(\bm n,\bm x,\bm v,t)
{\rm e}^{-\bm x\cdot\bm T_j/\lambda_j} \nonumber\\
&=& \sum_{j=1}^3\int\!{\rm d}\Omega
\sum_{\bm n} \left[2T_{js}n_s\Gamma_j(\bm n,t)+T_{js}^2\Gamma_j^*(\bm n,t)\right]
P(\bm n,\bm x,\bm v,t)
{\rm e}^{-\bm x\cdot\bm T_j/\lambda_j} \nonumber\\
&=& \sum_{j=1}^3 \left[2T_{js}\langle \tilde{n}_s\Gamma_j(\tilde{\bm n},t){\rm e}^{-\tilde{\bm x}\cdot\bm T_j/\lambda_j}\rangle
+ T_{js}^2\langle\Gamma_j^*(\tilde{\bm n},t){\rm e}^{-\tilde{\bm x}\cdot\bm T_j/\lambda_j}\rangle\right],
\end{eqnarray}

where $\Gamma_j^*(\bm n,t)=\Gamma_j^+(\bm n,t)+\Gamma_j^-(\bm n,t)$.

\begin{eqnarray}\label{eq_B2}
\frac{{\rm d}D_s}{{\rm d}t}
&=& \frac{{\rm d}(\langle \tilde{n}_s^2\rangle-\langle \tilde{n}_s\rangle^2)}{{\rm d}t}
= \frac{{\rm d}\langle \tilde{n}_s^2\rangle}{{\rm d}t}-2\langle \tilde{n}_s\rangle\frac{{\rm d}\langle \tilde{n}_s\rangle}{{\rm d}t} \nonumber\\
&=& \sum_{j=1}^3 \left[2T_{js}\langle \tilde{n}_s\Gamma_j(\tilde{\bm n},t){\rm e}^{-\tilde{\bm x}\cdot\bm T_j/\lambda_j}\rangle
+ T_{js}^2\langle\Gamma_j^*(\tilde{\bm n},t){\rm e}^{-\tilde{\bm x}\cdot\bm T_j/\lambda_j}\rangle \right]
- 2\sum_{j=1}^3 T_{js}\langle \tilde{n}_s\rangle\langle\Gamma_j(\tilde{\bm n},t){\rm e}^{-\tilde{\bm x}\cdot\bm T_j/\lambda_j}\rangle \nonumber\\
&=& \sum_{j=1}^3 \left\{2T_{js}{\rm {Cov}}\left[\Gamma_j(\tilde{\bm n},t){\rm e}^{-\tilde{\bm x}\cdot\bm T_j/\lambda_j},\tilde{n}_s \right]
+ T_{js}^2\langle\Gamma_j^*(\tilde{\bm n},t)
{\rm e}^{-\tilde{\bm x}\cdot\bm T_j/\lambda_j}\rangle \right\},
\end{eqnarray}

where ${\rm{Cov}}[\cdots,\cdots]$ denotes the covariance of two random variables, and we use ${\rm{Var}}[\cdots]$ to denote the variance of a variable in the following context.

\begin{eqnarray}\label{eq_B3}
&&\frac{{\rm d}\langle\tilde{n}_1\tilde{n}_2\rangle}{{\rm d}t}
= \sum_{j=1}^3\sum_\pm\int\!{\rm d}\Omega
\left[\sum_{\bm n}n_1 n_2\Gamma_j^\pm(\bm n\mp\bm T_j,t) P(\bm n\mp \bm T_j,\bm x,\bm v,t)
-\sum_{\bm n}n_1 n_2\Gamma_j^\pm(\bm n,t)P(\bm n,\bm x,\bm v,t)\right]
{\rm e}^{-\bm x\cdot\bm T_j/\lambda_j} \nonumber\\
&&= \sum_{j=1}^3\sum_\pm\int\!{\rm d}\Omega
\sum_{\bm n} \left[(n_1\mp T_{j1})(n_2\mp T_{j2})\pm T_{j1}(n_2\mp T_{j2})
\pm T_{j2}(n_1\mp T_{j1})+ T_{j1}T_{j2} \right]\cdot  \nonumber\\
&&\quad \Gamma_j^\pm(\bm n\mp \bm T_j,t)
P(\bm n\mp\bm T_j,\bm x,\bm v,t)
{\rm e}^{-\bm x\cdot\bm T_j/\lambda_j}
- \sum_{j=1}^3\sum_\pm\int\!{\rm d}\Omega
\sum_{\bm n}n_1 n_2\Gamma_j^\pm(\bm n,t)
P(\bm n,\bm x,\bm v,t)
{\rm e}^{-\bm x\cdot\bm T_j/\lambda_j} \nonumber\\
&&= \sum_{j=1}^3\sum_\pm\int\!{\rm d}\Omega
\sum_{\bm n}(T_{j1}T_{j2}\pm T_{j1}n_2 \pm T_{j2}n_1)
\Gamma_j^\pm(\bm n,t)
P(\bm n,\bm x,\bm v,t)
{\rm e}^{-\bm x\cdot\bm T_j/\lambda_j} \nonumber\\
&&= \sum_{j=1}^3 \left[\langle(T_{j1}\tilde{n}_2 + T_{j2}\tilde{n}_1)\Gamma_j(\tilde{\bm n},t)
{\rm e}^{-\tilde{\bm x}\cdot\bm T_j/\lambda_j}\rangle
+ T_{j1}T_{j2}\langle\Gamma_j^*(\tilde{\bm n},t){\rm e}^{-\tilde{\bm x}\cdot\bm T_j/\lambda_j}\rangle\right]. \nonumber\\
\end{eqnarray}

\begin{eqnarray}\label{eq_B4}
\frac{{\rm d}\Sigma_s}{{\rm d}t}
&=& \frac{{\rm d}(\langle\tilde{n}_1\tilde{n}_2\rangle
-\langle\tilde{n}_1\rangle \langle\tilde{n}_2\rangle)}{{\rm d}t}
= \frac{{\rm d}\langle\tilde{n}_1\tilde{n}_2\rangle}{{\rm d}t}
-\langle \tilde{n}_1\rangle\frac{{\rm d}\langle\tilde{n}_2\rangle}{{\rm d}t}
-\langle \tilde{n}_2\rangle\frac{{\rm d}\langle\tilde{n}_1\rangle}{{\rm d}t} \nonumber\\
&=& \sum_{j=1}^3 \left[\langle(T_{j1}\tilde{n}_2 + T_{j2}\tilde{n}_1)\Gamma_j(\tilde{\bm n},t)
{\rm e}^{-\tilde{\bm x}\cdot\bm T_j/\lambda_j}\rangle
+ T_{j1}T_{j2}\langle\Gamma_j^*(\tilde{\bm n},t)
{\rm e}^{-\tilde{\bm x}\cdot\bm T_j/\lambda_j}\rangle \right] \nonumber\\
&&\quad - \langle\tilde{n}_1\rangle \sum_{j=1}^3 T_{j2}
\langle\Gamma_j(\tilde{\bm n},t)
{\rm e}^{-\tilde{\bm x}\cdot\bm T_j/\lambda_j}\rangle
- \langle\tilde{n}_2\rangle \sum_{j=1}^3 T_{j1}
\langle\Gamma_j(\tilde{\bm n},t)
{\rm e}^{-\tilde{\bm x}\cdot\bm T_j/\lambda_j}\rangle \nonumber\\
&=& \sum_{j=1}^3 \left[\langle(T_{j1}\tilde{n}_2 + T_{j2}\tilde{n}_1)
\Gamma_j(\tilde{\bm n},t)
{\rm e}^{-\tilde{\bm x}\cdot\bm T_j/\lambda_j}\rangle
+ T_{j1}T_{j2}\langle\Gamma_j^*(\tilde{\bm n},t)
{\rm e}^{-\tilde{\bm x}\cdot\bm T_j/\lambda_j}\rangle \right] \nonumber\\
&&\quad - \sum_{j=1}^3 (T_{j1}\langle\tilde{n}_2\rangle + T_{j2}\langle\tilde{n}_1\rangle)
\langle\Gamma_j(\tilde{\bm n},t)
{\rm e}^{-\tilde{\bm x}\cdot\bm T_j/\lambda_j}\rangle \nonumber\\
&=& \sum_{j=1}^3 \! {\rm {Cov}}\left[\Gamma_j(\tilde{\bm n},t)
{\rm e}^{-\tilde{\bm x}\cdot\bm T_j/\lambda_j},T_{j1}\tilde{n}_2+T_{j2}\tilde{n}_1 \right]
- \langle\Gamma_2^*(\tilde{\bm n},t)
{\rm e}^{-\tilde{\bm x}\cdot\bm T_j/\lambda_j}\rangle. \nonumber\\
\end{eqnarray}

\begin{eqnarray}\label{eq_B5}
\frac{{\rm d}\langle\tilde{x}_s^2\rangle}{{\rm d}t}
&=& -\sum_{\bm n}\int\! x_s^2 v_s\frac{\partial P(\bm n,\bm x,\bm v,t)}{\partial x_s}{\rm d}\Omega
- \sum_{\bm n}\int\!
\frac{\partial \left[x_s^2(-\gamma_s v_s-\omega_s^2 x_s+F_s(\bm n,t)/m_s)P(\bm n,\bm x,\bm v,t) \right]}
{\partial v_s}{\rm d}\Omega \nonumber\\
&=& -\sum_{\bm n}\int\! v_s
\left\{\frac{\partial \left[x_s^2 P(\bm n,\bm x,\bm v,t)\right]}{\partial x_s}
- 2x_s P(\bm n,\bm x,\bm v,t) \right\}
{\rm d}\Omega \nonumber\\
&=& 2 \langle \tilde{x}_s\tilde{v}_s\rangle.
\end{eqnarray}

\begin{eqnarray}\label{eq_B6}
\frac{{\rm d}\langle\tilde{v}_s^2\rangle}{{\rm d}t}
&=& -\sum_{\bm n}\int\! v_s^3
\frac{\partial P(\bm n,\bm x,\bm v,t)}{\partial x_s}{\rm d}\Omega
- \sum_{\bm n}\int\! v_s^2
\frac{\partial \left[(-\gamma_s v_s-\omega_s^2 x_s+F_s(\bm n,t)/m_s)P(\bm n,\bm x,\bm v,t)\right]}
{\partial v_s}{\rm d}\Omega \nonumber\\
&=& -\sum_{\bm n}\int\! \frac{\partial \left[v_s^2(-\gamma_s v_s-\omega_s^2 x_s+F_s(\bm n,t)/m_s)
P(\bm n,\bm x,\bm v,t)\right]}{\partial v_s}{\rm d}\Omega \nonumber\\
&&\quad + \sum_{\bm n}\int\! 2v_s(-\gamma_s v_s-\omega_s^2 x_s+F_s(\bm n,t)/m_s)
P(\bm n,\bm x,\bm v,t)
{\rm d}\Omega \nonumber\\
&=& 2 \left[-\gamma_s\langle \tilde{v}_s^2\rangle - \omega_s^2 \langle \tilde{x}_s\tilde{v}_s\rangle +
\frac{\langle F_s(\tilde{\bm n},t)\tilde{v}_s\rangle}{m_s} \right].
\end{eqnarray}

\begin{eqnarray}\label{eq_B7}
\frac{{\rm d}\langle\tilde{x}_s\tilde{v}_s\rangle}{{\rm d}t}
&=& -\sum_{\bm n}\int\! x_s v_s^2
\frac{\partial P(\bm n,\bm x,\bm v,t)}{\partial x_s}{\rm d}\Omega
- \sum_{\bm n}\int\! x_s v_s
\frac{\partial \left[(-\gamma_s v_s-\omega_s^2 x_s+F_s(\bm n,t)/m_s)P(\bm n,\bm x,\bm v,t)\right]}
{\partial v_s}{\rm d}\Omega \nonumber\\
&=& -\sum_{\bm n}\int\! v_s^2
\left[\frac{\partial x_s P(\bm n,\bm x,\bm v,t)}{\partial x_s}
- P(\bm n,\bm x,\bm v,t)\right]{\rm d}\Omega \nonumber\\
&& \quad + \sum_{\bm n}\int\! x_s
\frac{\partial \left[v_s \left(-\gamma_s v_s-\omega_s^2 x_s+F_s(\bm n,t)/m_s \right)
P(\bm n,\bm x,\bm v,t) \right]} {\partial v_s}{\rm d}\Omega \nonumber\\
&& \quad + \sum_{\bm n}\int\! \left(-\gamma_s v_s-\omega_s^2 x_s+F_s(\bm n,t)/m_s \right)
P(\bm n,\bm x,\bm v,t) {\rm d}\Omega \nonumber\\
&=& \langle \tilde{v}_s^2\rangle-\gamma_s\langle \tilde{x}_s\tilde{v}_s\rangle - \omega_s^2\langle \tilde{x}_s^2\rangle +
\frac{\langle F_s(\tilde{\bm n},t)\tilde{x}_s\rangle}{m_s}.
\end{eqnarray}

\begin{eqnarray}\label{eq_B8}
\frac{{\rm d}{\rm Var}[\tilde{x}_s]}{{\rm d}t}
&=& \frac{{\rm d}(\langle \tilde{x}_s^2\rangle-\langle \tilde{x}_s\rangle^2)}{{\rm d}t}
= \frac{{\rm d}\langle \tilde{x}_s^2\rangle}{{\rm d}t}
- 2\langle \tilde{x}_s\rangle\frac{{\rm d}\langle \tilde{x}_s\rangle}{{\rm d}t} \nonumber\\
&=& 2\langle \tilde{x}_s \tilde{v}_s\rangle - 2\langle\tilde{x}_s\rangle \langle\tilde{v}_s\rangle
= 2{\rm{Cov}}[\tilde{x}_s,\tilde{v}_s].
\end{eqnarray}

\begin{eqnarray}\label{eq_B9}
\frac{{\rm d}{\rm Var}[\tilde{v}_s]}{{\rm d}t}
&=& \frac{{\rm d}(\langle \tilde{v}_s^2\rangle-\langle \tilde{v}_s\rangle^2)}{{\rm d}t}
= \frac{{\rm d}\langle \tilde{v}_s^2\rangle}{{\rm d}t}
- 2\langle \tilde{v}_s\rangle\frac{{\rm d}\langle \tilde{v}_s\rangle}{{\rm d}t} \nonumber\\
&=& 2 \left[-\gamma_s\langle \tilde{v}_s ^2\rangle
- \omega_s^2\langle\tilde{x}_s\tilde{v}_s\rangle
+ \frac{\langle F_s(\tilde{\bm n},t)\tilde{v}_s\rangle}{m_s} \right]
- 2\langle\tilde{v}_s\rangle \left[-\gamma_s\langle \tilde{v}_s\rangle
- \omega_s^2\langle\tilde{x}_s\rangle
+ \frac{\langle F_s(\tilde{\bm n},t)\rangle}{m_s} \right] \nonumber\\
&=& 2 \left\{-\gamma_s{\rm{Var}}[\tilde{v}_s]-\omega_s^2{\rm{Cov}}[\tilde{x}_s,\tilde{v}_s]
+ \frac{\langle F_s(\tilde{\bm n},t)(\tilde{v}_s-\langle\tilde{v}_s\rangle)\rangle}{m_s} \right\}.
\end{eqnarray}

\begin{eqnarray}\label{eq_B10}
\frac{{\rm d}{\rm Cov}[\tilde{x}_s,\tilde{v}_s]}{{\rm d}t}
&=& \frac{{\rm d}(\langle\tilde{x}_s\tilde{v}_s\rangle
- \langle\tilde{x}_s\rangle \langle\tilde{v}_s\rangle)}{{\rm d}t}
= \frac{{\rm d}\langle\tilde{x}_s\tilde{v}_s\rangle}{{\rm d}t}
- \langle\tilde{x}_s\rangle\frac{{\rm d}\langle\tilde{v}_s\rangle}{{\rm d}t}
- \langle\tilde{v}_s\rangle\frac{{\rm d}\langle\tilde{x}_s\rangle}{{\rm d}t} \nonumber\\
&=& \langle\tilde{v}_s^2\rangle-\gamma_s\langle\tilde{x}_s\tilde{v}_s\rangle
- \omega_s^2\langle\tilde{x}_s^2\rangle
+ \frac{\langle F_s(\tilde{\bm n},t)\tilde{x}_s\rangle\rangle}{m_s}
- \langle\tilde{x}_s\rangle \left[-\gamma_s\langle\tilde{v}_s\rangle
- \omega_s^2\langle\tilde{x}_s\rangle + \frac{\langle F_s(\tilde{\bm n},t)\rangle}{m_s} \right]
- \langle\tilde{v}_s\rangle^2 \nonumber\\
&=& {\rm{Var}}[\tilde{v}_s]-\gamma_s{\rm{Cov}}[\tilde{x}_s,\tilde{v}_s]
-\omega_s^2{\rm{Var}}[\tilde{x}_s]
+ \frac{\langle F_s(\tilde{\bm n},t)(\tilde{x}_s-\langle\tilde{x}_s\rangle)\rangle}{m_s}.
\end{eqnarray}

\begin{eqnarray}\label{eq_B11}
\frac{{\rm d}\langle\tilde{x}_1\tilde{x}_2\rangle}{{\rm d}t}
&=& -\sum_{\bm n}\int\! \left[x_1 x_2 v_1\frac{\partial P(\bm n,\bm x,\bm v,t)}{\partial x_1}
+ x_1 x_2 v_2\frac{\partial P(\bm n,\bm x,\bm v,t)}{\partial x_2}\right]
{\rm d}\Omega \nonumber\\
&=& -\sum_{\bm n}\int\! v_1 x_2
\left[\frac{\partial [x_1 P(\bm n,\bm x,\bm v,t)]}{\partial x_1}- P(\bm n,\bm x,\bm v,t)\right]{\rm d}\Omega
- \sum_{\bm n}\int\! v_2 x_1 \left[\frac{\partial [x_2 P(\bm n,\bm x,\bm v,t)]}{\partial x_2}
- P(\bm n,\bm x,\bm v,t)\right]{\rm d}\Omega \nonumber\\
&=& \langle\tilde{x}_1\tilde{v}_2\rangle+\langle\tilde{x}_2\tilde{v}_1\rangle.
\end{eqnarray}

\begin{eqnarray}\label{eq_B12}
\frac{{\rm d}\langle\tilde{x}_{s'}\tilde{v}_s\rangle}{{\rm d}t}
&=& -\sum_{\bm n}\int\! x_{s'} v_s^2\frac{\partial P(\bm n,\bm x,\bm v,t)}{\partial x_s}
{\rm d}\Omega
- \sum_{\bm n}\int\! x_{s'} v_s
\frac{\partial \left[(-\gamma_s v_s-\omega_s^2 x_s+F_s(\bm n,t)/m_s)P(\bm n,\bm x,\bm v,t)\right]}
{\partial v_s}{\rm d}\Omega \nonumber\\
&& \quad - \sum_{\bm n}\int\! x_{s'} v_s v_{s'}\frac{\partial P(\bm n,\bm x,\bm v,t)}{\partial x_{s'}} {\rm d}\Omega
- \sum_{\bm n}\int\! x_{s'} v_s
\frac{\partial \left[(-\gamma_{s'} v_{s'}-\omega_{s'}^2 x_{s'} + F_{s'}(\bm n,t)/m_{s'})P(\bm n,\bm x,\bm v,t)\right]}
{\partial v_{s'}}{\rm d}\Omega \nonumber\\
&=& -\sum_{\bm n}\int\! v_s v_{s'}
\left[\frac{\partial [x_s P(\bm n,\bm x,\bm v,t)]}{\partial x_{s'}}- P(\bm n,\bm x,\bm v,t)\right]{\rm d}\Omega \nonumber\\
&& \quad + \sum_{\bm n}\int\! x_{s'}
\frac{\partial \left[v_s(-\gamma_s v_s-\omega_s^2 x_s+F_s(\bm n,t)/m_s)P(\bm n,\bm x,\bm v,t)\right]}
{\partial v_s}{\rm d}\Omega \nonumber\\
&& \quad + \sum_{\bm n}\int\! (-\gamma_s v_s-\omega_s^2 x_s+F_s(\bm n,t)/m_s)
P(\bm n,\bm x,\bm v,t)
{\rm d}\Omega \nonumber\\
&=& \langle \tilde{v}_s \tilde{v}_{s'}\rangle-\gamma_s\langle \tilde{x}_{s'}\tilde{v}_s\rangle - \omega_s^2\langle \tilde{x}_s \tilde{x}_{s'}\rangle + \frac{\langle F_s(\tilde{\bm n},t)\tilde{x}_{s'}\rangle}{m_s}.
\end{eqnarray}

\begin{eqnarray}\label{eq_B13}
\frac{{\rm d}\langle\tilde{v}_{s'}\tilde{v}_s\rangle}{{\rm d}t}
&=& -\sum_{\bm n}\int\! v_{s'} v_s^2\frac{\partial P(\bm n,\bm x,\bm v,t)}{\partial x_s}
{\rm d}\Omega
- \sum_{\bm n}\int\! v_{s'} v_s
\frac{\partial \left[(-\gamma_s v_s-\omega_s^2 x_s+F_s(\bm n,t)/m_s)P(\bm n,\bm x,\bm v,t)\right]}
{\partial v_s}{\rm d}\Omega \nonumber\\
&& \quad - \sum_{\bm n}\int\! v_s v_{s'}^2\frac{\partial P(\bm n,\bm x,\bm v,t)}{\partial x_{s'}}{\rm d}\Omega
- \sum_{\bm n}\int\! v_s v_{s'}\frac{\partial \left[(-\gamma_{s'} v_{s'}-\omega_{s'}^2 x_{s'}
+ F_{s'}(\bm n,t)/m_{s'})P(\bm n,\bm x,\bm v,t)\right]}{\partial v_{s'}} {\rm d}\Omega \nonumber\\
&=& -\sum_{\bm n}\int\! v_{s'} \frac{\partial \left[v_s(-\gamma_s v_s-\omega_s^2 x_s
+F_s(\bm n,t)/m_s) P(\bm n,\bm x,\bm v,t)\right]}{\partial v_s} \nonumber\\
&& \quad + \sum_{\bm n}\int\! v_{s'}(-\gamma_s v_s-\omega_s^2 x_s+F_s(\bm n,t)/m_s)
P(\bm n,\bm x,\bm v,t)]{\rm d}\Omega \nonumber\\
&& \quad - \sum_{\bm n}\int\! v_{s}\frac{\partial \left[v_{s'}(-\gamma_{s'} v_{s'}
- \omega_{s'}^2 x_{s'}+F_{s'}(\bm n,t)/m_{s'}) P(\bm n,\bm x,\bm v,t)\right]}
{\partial v_{s'}}{\rm d}\Omega \nonumber\\
&& \quad + \sum_{\bm n}\int\! v_s(-\gamma_{s'} v_{s'}-\omega_{s'}^2 x_{s'}+F_{s'}(\bm n,t)/m_{s'})
P(\bm n,\bm x,\bm v,t)
{\rm d}\Omega \nonumber\\
&=& -(\gamma_s+\gamma_{s'})\langle \tilde{v}_s \tilde{v}_{s'}\rangle
- \omega_s^2\langle \tilde{x}_s\tilde{v}_{s'}\rangle
- \omega_{s'}^2\langle \tilde{x}_{s'}\tilde{v}_s\rangle
+ \frac{\langle F_s(\tilde{\bm n},t)\tilde{v}_{s'}\rangle}{m_s}
+ \frac{\langle F_{s'}(\tilde{\bm n},t)\tilde{v}_s\rangle}{m_{s'}}.
\end{eqnarray}

\begin{eqnarray}\label{eq_B14}
\frac{{\rm d}{\rm{Cov}}[\tilde{x}_1,\tilde{x}_2]}{{\rm d}t}
&=& \frac{{\rm d}\langle\tilde{x}_1\tilde{x}_2\rangle}{{\rm d}t}
- \langle\tilde{x}_1\rangle\frac{{\rm d}\langle\tilde{x}_2\rangle}{{\rm d}t}
- \langle\tilde{x}_2\rangle\frac{{\rm d}\langle\tilde{x}_1\rangle}{{\rm d}t} \nonumber\\
&=& \langle\tilde{x}_1\tilde{v}_2\rangle+\langle\tilde{x}_2\tilde{v}_1\rangle
- \langle\tilde{x}_1\rangle \langle\tilde{v}_2\rangle
- \langle\tilde{x}_2\rangle \langle\tilde{v}_1\rangle \nonumber\\
&=& {\rm{Cov}}[\tilde{x}_1,\tilde{v}_2]+{\rm{Cov}}[\tilde{x}_2,\tilde{v}_1].
\end{eqnarray}

\begin{eqnarray}\label{eq_B15}
\frac{{\rm d}{\rm{Cov}}[\tilde{v}_1,\tilde{v}_2]}{{\rm d}t}
&=& \frac{{\rm d}\langle\tilde{v}_1\tilde{v}_2\rangle}{{\rm d}t}
- \langle\tilde{v}_1\rangle\frac{{\rm d}\langle\tilde{v}_2\rangle}{{\rm d}t}
- \langle\tilde{v}_2\rangle\frac{{\rm d}\langle\tilde{v}_1\rangle}{{\rm d}t} \nonumber\\
&=& -(\gamma_1+\gamma_2){\rm{Cov}}[\tilde{v}_1,\tilde{v}_2]
-\omega_1^2{\rm{Cov}}[\tilde{x}_1,\tilde{v}_2]
-\omega_2^2{\rm{Cov}}[\tilde{x}_2,\tilde{v}_1] \nonumber\\
&& \quad + \langle\frac{F_1(\bm n,t)(\tilde{v}_2-\langle\tilde{v}_2\rangle)}{m_1}\rangle
+ \langle\frac{F_2(\bm n,t)(\tilde{v}_1-\langle\tilde{v}_1\rangle)}{m_2}\rangle.
\end{eqnarray}

\begin{eqnarray}\label{eq_B16}
\frac{{\rm d}{\rm{Cov}}[\tilde{x}_{s'},\tilde{v}_s]}{{\rm d}t}
&=& \frac{{\rm d}(\langle\tilde{x}_{s'}\tilde{v}_s\rangle
- \langle\tilde{x}_{s'}\rangle \langle\tilde{v}_s\rangle)}{{\rm d}t} \nonumber\\
&=& \frac{{\rm d}\langle\tilde{x}_{s'}\tilde{v}_s\rangle}{{\rm d}t}
- \langle\tilde{x}_{s'}\rangle\frac{{\rm d}\langle\tilde{v}_s\rangle}{{\rm d}t}
- \langle\tilde{v}_s\rangle\frac{{\rm d}\langle\tilde{x}_{s'}\rangle}{{\rm d}t} \nonumber\\
&=& \langle\tilde{v}_s\tilde{v}_{s'}\rangle
-\gamma_s\langle\tilde{x}_{s'}\tilde{v}_s\rangle
-\omega_s^2\langle\tilde{x}_{s'}\tilde{x}_s\rangle
+\frac{\langle F_s(\tilde{\bm n},t)\tilde{x}_{s'}\rangle}{m_s}
- \langle\tilde{x}_{s'}\rangle
[-\gamma_s\langle\tilde{v}_s\rangle
-\omega_s^2\langle\tilde{x}_s\rangle
+\frac{\langle F_s(\tilde{\bm n},t)\rangle}{m_s}] \nonumber\\
&=& {\rm{Cov}}[\tilde{v}_s,\tilde{v}_{s'}]
-\gamma_s{\rm{Cov}}[\tilde{x}_{s'},\tilde{v}_{s}]
-\omega_s^2{\rm{Cov}}[\tilde{x}_s,\tilde{x}_{s'}]
+ \frac{\langle F_{s}(\tilde{\bm n},t)(\tilde{x}_{s'}-\langle\tilde{x}_{s'}\rangle)\rangle}{m_{s}}.
\end{eqnarray}

\begin{eqnarray}\label{eq_F1}
\frac{{\rm d}\langle\tilde{x}_s\tilde{n}_{s'}\rangle}{{\rm d}t}
&=& \sum_{j=1}^3\int\!
\sum_{\bm n, \pm} \left\{ \left[(n_{s'}\mp T_{js'})\pm T_{js'}\right]
\Gamma_j^\pm(\bm n\mp\bm T_j,t)P(\bm n\mp\bm T_j,\bm x,\bm v,t)
- n_{s'}\Gamma_j^\pm(\bm n,t)P(\bm n,\bm x,\bm v,t)\right\}
x_s{\rm e}^{-\bm x\cdot\bm T_j/\lambda_j}{\rm d}\Omega \nonumber\\
&& \quad - \sum_{\bm n}\int\!
n_{s'}x_s v_s\frac{\partial P(\bm n,\bm x,\bm v,t)}{\partial x_s}{\rm d}\Omega
- \sum_{\bm n}\int\! n_{s'}x_s
\frac{\partial \left[(-\gamma_s v_s-\omega_s^2 x_s+F_s(\bm n,t)/m_s)P(\bm n,\bm x,\bm v,t)\right]}
{\partial v_s}{\rm d}\Omega \nonumber\\
&=& \sum_{j=1}^3\int\!\sum_{\bm n, \pm}\pm T_{js'}\Gamma_j^\pm(\bm n,t)
P(\bm n,\bm x,\bm v,t)
x_s{\rm e}^{-\bm x\cdot\bm T_j/\lambda_j}{\rm d}\Omega
- \sum_{\bm n}\int\!
n_{s'}v_s \left[\frac{\partial x_s P(\bm n,\bm x,\bm v,t)}{\partial x_s}-P(\bm n,\bm x,\bm v,t) \right]
{\rm d}\Omega \nonumber\\
&=& \sum_{j=1}^3 T_{js'}\langle\tilde{x}_s\Gamma_j(\tilde{\bm n},t)
{\rm e}^{-\tilde{\bm x}\cdot\bm T_j/\lambda_j}\rangle
+\langle\tilde{v}_s\tilde{n}_{s'}\rangle.
\end{eqnarray}

\begin{eqnarray}\label{eq_F2}
\frac{{\rm d}\langle\tilde{v}_s\tilde{n}_{s'}\rangle}{{\rm d}t}
&=& \sum_{j=1}^3\int\!
\sum_{\bm n, \pm} \left\{\left[(n_{s'}\mp T_{js'})\pm T_{js'}\right]
\Gamma_j^\pm(\bm n\mp\bm T_j,t)P(\bm n\mp\bm T_j,\bm x,\bm v,t)
- n_{s'}\Gamma_j^\pm(\bm n,t)P(\bm n,\bm x,\bm v,t)\right\}
v_s{\rm e}^{-\bm x\cdot\bm T_j/\lambda_j}{\rm d}\Omega \nonumber\\
&& \quad - \sum_{\bm n}\int\!
n_{s'}v_s^2\frac{\partial P(\bm n,\bm x,\bm v,t)}{\partial x_s}{\rm d}\Omega
- \sum_{\bm n}\int\! n_{s'}v_s
\frac{\partial \left[(-\gamma_s v_s-\omega_s^2 x_s+F_s(\bm n,t)/m_s)P(\bm n,\bm x,\bm v,t)\right]}
{\partial v_s}{\rm d}\Omega \nonumber\\
&=& \sum_{j=1}^3\sum_\pm\int\! \sum_{\bm n}\pm T_{js'}\Gamma_j^\pm(\bm n,t)P(\bm n,\bm x,\bm v,t)
v_s{\rm e}^{-\bm x\cdot\bm T_j/\lambda_j}{\rm d}\Omega \nonumber\\
&& \quad - \sum_{\bm n}\int\!
n_{s'}\frac{\partial \left[v_s(-\gamma_s v_s-\omega_s^2 x_s+F_s(\bm n,t)/m_s)P(\bm n,\bm x,\bm v,t)\right]}
{\partial v_s}{\rm d}\Omega \nonumber\\
&& \quad - \sum_{\bm n}\int\!n_{s'}
(-\gamma_s v_s-\omega_s^2 x_s+F_s(\bm n,t)/m_s)P(\bm n,\bm x,\bm v,t)
{\rm d}\Omega \nonumber\\
&=& \sum_{j=1}^3 T_{js'}\langle\tilde{v}_s\Gamma_j(\tilde{\bm n},t)
{\rm e}^{-\tilde{\bm x}\cdot\bm T_j/\lambda_j}\rangle
- \gamma_s\langle\tilde{v}_s\tilde{n}_{s'}\rangle
- \omega_s^2\langle\tilde{x}_s\tilde{n}_{s'}\rangle
+ \frac{\langle F_s(\tilde{\bm n},t)\tilde{n}_{s'}\rangle}{m_s}.
\end{eqnarray}

\begin{eqnarray}\label{eq_F3}
\frac{{\rm d}{\rm{Cov}}\left[\tilde{x}_s, \tilde{n}_{s'}\right]}{{\rm d}t}
&=& \frac{{\rm d}\langle\tilde{x}_{s}\tilde{n}_{s'}\rangle}{{\rm d}t}
- \langle\tilde{x}_s\rangle\frac{{\rm d}\langle\tilde{n}_{s'}\rangle}{{\rm d}t} - \langle\tilde{n}_{s'}\rangle\frac{{\rm d}\langle\tilde{x}_{s}\rangle}{{\rm d}t} \nonumber\\
&=& \sum_{j=1}^3 T_{js'}\langle\tilde{x}_s\Gamma_j(\tilde{\bm n},t)
{\rm e}^{-\tilde{\bm x}\cdot\bm T_j/\lambda_j}\rangle
+ \langle\tilde{v}_s\tilde{n}_{s'}\rangle
- \langle\tilde{x}_s\rangle
\sum_{j=1}^3 T_{js'}\langle\Gamma_j(\tilde{\bm n},t)
{\rm e}^{-\tilde{\bm x}\cdot\bm T_j/\lambda_j}\rangle
- \langle\tilde{v}_s\rangle \langle\tilde{n}_{s'}\rangle \nonumber\\
&=& \sum_{j=1}^3 T_{js'}\langle(\tilde{x}_s-\langle\tilde{x}_s\rangle)
\Gamma_j(\tilde{\bm n},t)
{\rm e}^{-\tilde{\bm x}\cdot\bm T_j/\lambda_j}\rangle
+ {\rm {Cov}}\left[\tilde{v}_s, \tilde{n}_{s'}\right] \nonumber\\
&=& {\rm Cov} \left[\tilde{x}_s(t), \Gamma_j(\tilde{\bm n},t){\rm e}^{-\tilde{\bm x}\cdot\bm T_j/\lambda_j} \right]
+ {\rm {Cov}}\left[\tilde{v}_s, \tilde{n}_{s'}\right].
\end{eqnarray}

\begin{eqnarray}\label{eq_F4}
\frac{{\rm d}{\rm{Cov}}\left[\tilde{v}_s, \tilde{n}_{s'}\right]}{{\rm d}t}
&=& \frac{{\rm d}\langle\tilde{v}_{s}\tilde{n}_{s'}\rangle}{{\rm d}t}
- \langle\tilde{v}_s\rangle\frac{{\rm d}\langle\tilde{n}_{s'}\rangle}{{\rm d}t} - \langle\tilde{n}_{s'}\rangle\frac{{\rm d}\langle\tilde{v}_{s}\rangle}{{\rm d}t} \nonumber\\
&=& \sum_{j=1}^3 T_{js'}\langle\tilde{v}_s\Gamma_j(\tilde{\bm n},t)
{\rm e}^{-\tilde{\bm x}\cdot\bm T_j/\lambda_j}\rangle
- \gamma_s\langle\tilde{v}_s\tilde{n}_{s'}\rangle
- \omega_s^2\langle\tilde{x}_s\tilde{n}_{s'}\rangle
+ \frac{\langle F_s(\tilde{\bm n},t)\tilde{n}_{s'}\rangle}{m_s} \nonumber\\
&& \quad - \langle\tilde{v}_s\rangle
\sum_{j=1}^3 T_{js'}\langle\Gamma_j(\tilde{\bm n},t)
{\rm e}^{-\tilde{\bm x}\cdot\bm T_j/\lambda_j}\rangle
- \langle\tilde{n}_{s'}\rangle
\left[-\gamma_s \langle\tilde{v}_s\rangle-\omega_s^2 \langle\tilde{x}_s\rangle
+ \frac{\langle F_s(\tilde{\bm n},t)\rangle}{m_s} \right] \nonumber\\
&=& \sum_{j=1}^3 T_{js'}
{\rm Cov} \left[\tilde{v}_s(t), \Gamma_j(\tilde{\bm n},t){\rm e}^{-\tilde{\bm x}\cdot\bm T_j/\lambda_j} \right]
+ \frac{\langle F_s(\tilde{\bm n},t)
(\tilde{n}_{s'}-\langle\tilde{n}_{s'}\rangle)\rangle}{m_s} \nonumber\\
&& \quad - \gamma_s {\rm{Cov}}\left[\tilde{v}_s, \tilde{n}_{s'}\right]
- \omega_s^2 {\rm{Cov}}\left[\tilde{x}_s, \tilde{n}_{s'}\right].
\end{eqnarray}

\section{Covariance of Complicated Formulations}\label{App_C}
Assume ${\tilde x}_1$,${\tilde x}_2$ and ${\tilde n}_1$,${\tilde n}_2$ have the multivariate normal distribution, $\Sigma_1={\rm Cov}[{\tilde x}_1,{\tilde n}_1]$, $\Sigma_2={\rm Cov}[{\tilde x}_2,{\tilde n}_1]$, we can obtain the following rules to formulate the covariance based on the normal distribution:
\begin{equation}\label{eq_C1}
\left\langle({\tilde x}_1-\langle {\tilde x}_1 \rangle) ({\tilde n}_1-\langle {\tilde n}_1 \rangle)^{l_1} \right\rangle
= l_1 \Sigma_1 \left\langle({\tilde n}_1-\langle {\tilde n}_1 \rangle)^{l_1-1} \right\rangle,
\end{equation}
\begin{equation}\label{eq_C2}
\left\langle ({\tilde n}_1- \langle {\tilde n}_1 \rangle) ({\tilde x}_1- \langle {\tilde x}_1 \rangle)^{l_1} ({\tilde x}_1- \langle {\tilde x}_2 \rangle)^{l_2} \right\rangle
= l_1 \Sigma_1 \left\langle({\tilde x}_1- \langle {\tilde x}_1 \rangle)^{l_1-1} ({\tilde x}_1-\langle {\tilde x}_2 \rangle)^{l_2} \right\rangle
+l_2 \Sigma_2 \left\langle({\tilde x}_1- \langle {\tilde x}_1 \rangle)^{l_1} ({\tilde x}_1- \langle {\tilde x}_2 \rangle)^{l_2-1} \right\rangle,
\end{equation}
Using $f$ to denote a function, we hence have more general rules by adopting Taylor expansions:
\begin{eqnarray}\label{eq_C3}
{\rm Cov}\left[{\tilde n}_1,f({\tilde x}_1 )\right]
= \left\langle({\tilde n}_1-\langle {\tilde n}_1 \rangle)f({\tilde x}_1 )\right\rangle
&=& \sum_{l_1=1}^{\infty} \frac{f^{(l_1 )}(\langle {\tilde x}_1 \rangle)}{l_1 !}
\left\langle ({\tilde n}_1-\langle {\tilde n}_1 \rangle) ({\tilde x}_1-\langle {\tilde x}_1 \rangle)^{l_1} \right\rangle  \nonumber\\
&=& \sum_{l_1=1}^{\infty} \frac{f^{(l_1 )}(\langle {\tilde x}_1 \rangle)}{l_1 !}  l_1
\left\langle ({\tilde x}_1- \langle {\tilde x}_1 \rangle)^{l_1-1} \right\rangle \Sigma_1 \nonumber\\
&=&\Sigma_1 \sum_{l_1=0}^{\infty} \left\langle( \frac{f^{(l_1+1)}(\langle {\tilde x}_1 \rangle))}{l_1 !}
 ({\tilde x}_1-\langle {\tilde x}_1 \rangle)^{l_1} \right\rangle  \nonumber\\
&=&\Sigma_1 \langle f' ({\tilde x}_1 )\rangle,
\end{eqnarray}
\begin{eqnarray}\label{eq_C4}
{\rm Cov} \left[{\tilde n}_1,f({\tilde x}_1,{\tilde x}_2 )\right]
&=& \left\langle({\tilde n}_1- \langle {\tilde n}_1 \rangle)f({\tilde x}_1,{\tilde x}_2 )\right\rangle \nonumber\\
&=& \sum_{l_1, l_2}^{\infty} \left\langle \frac{f^{(l_1,l_2)} (\langle {\tilde x}_1 \rangle, \langle {\tilde x}_2 \rangle)}{l_1 !l_2 !}
 ({\tilde n}_1-\langle {\tilde n}_1 \rangle) ({\tilde x}_1-\langle {\tilde x}_1 \rangle)^{l_1} ({\tilde x}_2-\langle {\tilde x}_2 \rangle)^{l_2} \right\rangle \nonumber\\
&=& \sum_{l_1, l_2}^{\infty} \frac{f^{(l_1,l_2)} (\langle {\tilde x}_1 \rangle, \langle {\tilde x}_2 \rangle)}{l_1 !l_2 !}
\left[l_1 \Sigma_1 \left\langle ({\tilde x}_1- \langle {\tilde x}_1 \rangle)^{l_1-1} ({\tilde x}_2- \langle {\tilde x}_2 \rangle)^{l_2}\right\rangle
+l_2 \Sigma_2 \left\langle ({\tilde x}_1-\langle {\tilde x}_1 \rangle)^{l_1} ({\tilde x}_2-\langle {\tilde x}_2 \rangle)^{l_2-1} \right\rangle \right] \nonumber\\
&=& \Sigma_1 \left\langle \frac{\partial f({\tilde x}_1,{\tilde x}_2)}{\partial {\tilde x}_1} \right\rangle
+ \Sigma_2 \left\langle \frac{\partial f({\tilde x}_1,{\tilde x}_2)}{\partial {\tilde x}_2} \right\rangle.
\end{eqnarray}

For calculation of ${\rm Cov}\left[v_s, \Gamma_j (\tilde {\bm n},\tilde {\bm x},t)\right]$
we can assume $\Sigma_s={\rm Cov}[{\tilde x}_s,{\tilde v}_s]$, $X_s={\rm Cov}[{\tilde x}_s,{\tilde n}_s]$, and $V_s={\rm Cov}[{\tilde v}_s,{\tilde n}_s]$. Then using the rule (\ref{eq_C4}) with the relation $\Gamma_j (\tilde {\bm n},\tilde {\bm x},t) = K_j (\tilde {\bm x}) \Gamma_j (\tilde {\bm n},t)$, we have
\begin{eqnarray}\label{eq_C5}
{\rm Cov}\left[v_s, \Gamma_j (\tilde {\bm n},\tilde {\bm x},t)\right]
\cong \Sigma_s \left\langle \frac{\partial K_j(\tilde {\bm x})}{\partial {\tilde x}_s} \Gamma_j (\tilde {\bm n},t)  \right\rangle
+V_s \left\langle K_j(\tilde {\bm x}) \frac{\partial \Gamma_j (\tilde {\bm n},t)}{\partial {\tilde x}_s}  \right\rangle,
\end{eqnarray}
In the domain of $| \langle U_j^{\pm} (n,t)\rangle | \gg k_B T$, $ \Gamma_j (\tilde {\bm n},t)$ can be well approximated by a linear function of $\tilde {\bm n}$. Thus,
\begin{eqnarray}\label{eq_C6}
\left\langle K_j(\tilde {\bm x}) \frac{\partial \Gamma_j (\tilde {\bm n},t)}{\partial {\tilde x}_s}  \right\rangle
\cong \left\langle \frac{\partial \Gamma_j (\tilde {\bm n},t))}{\partial {\tilde n}_s} \right\rangle
 \left\langle K_j (\tilde {\bm x})\right\rangle,
\end{eqnarray}
\begin{eqnarray}\label{eq_C7}
\left\langle \frac{\partial K_j(\tilde {\bm x})}{\partial {\tilde x}_s} \Gamma_j (\tilde {\bm n},t)  \right\rangle
\cong \left\langle \Gamma_j (\tilde {\bm n},t) \right\rangle
\left\langle\frac{\partial K_j(\tilde {\bm x})}{\partial {\tilde x}_s} \right\rangle
+X_1 \left\langle \frac{\partial \Gamma_j (\tilde {\bm n},t)}{{\partial \tilde n}_1} \right\rangle
\left\langle \frac{\partial^2 K_j(\tilde {\bm x})}{\partial {\tilde x}_s \partial {\tilde x}_1} \right\rangle
+X_2 \left\langle \frac{\partial \Gamma_j (\tilde {\bm n},t)}{\partial {\tilde n}_2} \right\rangle
\left\langle \frac{\partial^2 K_j(\tilde {\bm x})}{\partial {\tilde x}_s \partial {\tilde x}_2} \right\rangle. \quad \!\!
\end{eqnarray}
In above,
\begin{eqnarray}\label{eq_C8}
\left\langle\frac{\partial K_j(\tilde {\bm x})}{\partial {\tilde x}_s} \right\rangle
=-\frac{T_{js}}{\lambda_j}  \left\langle K_j(\tilde {\bm x})\right\rangle
\end{eqnarray}
\begin{eqnarray}\label{eq_C9}
\left\langle \frac{\partial^2 K_j(\tilde {\bm x})}{\partial {\tilde x}_s \partial {\tilde x}_s'} \right\rangle
=\frac{T_{js}}{\lambda_j} \frac{T_{js'}}{\lambda_j}  \left\langle K_j(\tilde {\bm x})\right\rangle
\end{eqnarray}
We hence obtain
\begin{eqnarray}\label{eq_C10}
&&{\rm Cov}\left[x_s, \Gamma_j (\tilde {\bm n},\tilde {\bm x},t)\right] \\
&&\cong \langle K_j(\tilde{\bm x})\rangle \left\{
V_s \left\langle \frac{\partial \Gamma_j (\tilde {\bm n},t)}{{\partial \tilde n}_s} \right\rangle
-\Sigma_s \left\langle \Gamma_j (\tilde {\bm n},t) \right\rangle \left(\frac{T_{js}}{\lambda_j} \right)
+\Sigma_s \left(\frac{T_{js}}{\lambda_j^2} \right) \left[
 X_1 \left\langle \frac{\partial \Gamma_j (\tilde {\bm n},t)}{{\partial \tilde n}_1} \right\rangle T_{j1}
+X_2 \left\langle \frac{\partial \Gamma_j (\tilde {\bm n},t)}{\partial {\tilde n}_2} \right\rangle T_{j2}
\right]\right\}. \nonumber
\end{eqnarray}

Additionally, we can simplify the formulation by defining
\begin{eqnarray}\label{eq_C11}
\bm g_j(t) = \langle\partial\Gamma_j(\tilde{\bm n},t)/\partial \tilde{\bm n}\rangle
\end{eqnarray}
as a vector ($g_{js}(t)$ as its $s$\textsuperscript{th} component) and defining
\begin{eqnarray}\label{eq_C12}
G_j(t)=\langle\Gamma_j(\tilde{\bm n},t)\rangle -\bm g_j(t){\rm diag}(\bm T_j)\bm X^{\rm T}/\lambda_j,
\end{eqnarray}
in which $\bm X=[X_1,X_2]$ is a vector and ${\rm diag}(\bm T_j)$ denotes the diagonal matrix with the diagonal $\bm T_j$. Note that $G_j$ is linear to $X_1$ and $X_2$. Thus, we obtain
\begin{eqnarray}\label{eq_C13}
{\rm Cov}\left[x_s, \Gamma_j (\tilde {\bm n},\tilde {\bm x},t)\right]
\cong \langle K_j(\tilde{\bm x})\rangle
\left[g_{js}(t)V_s-\frac{T_{js}}{\lambda_j}G_j (t)\Sigma_{ss}\right].
\end{eqnarray}

Similarly, we can derive the following covariance with the additional assumption $\Lambda_{ss}={\rm Var}[{{\tilde x}_s}]$.
\begin{eqnarray}\label{eq_C14}
{\rm Cov}\left[x_s, \Gamma_j (\tilde {\bm n},\tilde {\bm x},t)\right]
&&\cong \Lambda_s \left\langle \frac{\partial K_j(\tilde {\bm x})}{\partial {\tilde x}_s} \Gamma_j (\tilde {\bm n},t)  \right\rangle
+X_s \left\langle K_j(\tilde {\bm x}) \frac{\partial \Gamma_j (\tilde {\bm n},t)}{\partial {\tilde x}_s}  \right\rangle  \nonumber\\
&&\cong \langle K_j(\tilde{\bm x})\rangle
\left[g_{js}(t)X_s-\frac{T_{js}}{\lambda_j}G_j (t)\Lambda_{ss}\right].
\end{eqnarray}

In addition, we can formulate the covariance of $K_j (\tilde {\bm x})$ and $\Gamma_j (\tilde {\bm n},t)$, as an add-on term of $\langle\Gamma_j (\tilde {\bm n},\tilde {\bm x},t)\rangle$ in (\ref{eq_21}).
\begin{eqnarray}\label{eq_C15}
\langle\Gamma_j (\tilde {\bm n},\tilde {\bm x},t)\rangle
- \langle K_j (\tilde {\bm x})\rangle \langle \Gamma_j (\tilde {\bm n},t)\rangle
&&= {\rm Cov}\left[ K_j (\tilde {\bm x}), \Gamma_j (\tilde {\bm n},t)\right] \nonumber\\
&&\cong \left\langle \frac{\partial \Gamma_j (\tilde {\bm n},t)}{\partial {\tilde n}_1} \right\rangle
\left\langle \frac{\partial K_j(\tilde{\bm x})}{\partial {\tilde x}_1} \right\rangle X_1
+ \left\langle \frac{\partial \Gamma_j (\tilde {\bm n},t)}{\partial {\tilde n}_2} \right\rangle
\left\langle \frac{\partial K_j(\tilde{\bm x})}{\partial {\tilde x}_2} \right\rangle X_2 \nonumber\\
&&= -\left[ \left\langle \frac{\partial \Gamma_j (\tilde {\bm n},t)}{\partial {\tilde n}_1} \right\rangle
\frac{T_{j1}}{\lambda_j} X_1
+ \left\langle \frac{\partial \Gamma_j (\tilde {\bm n},t)}{\partial {\tilde n}_2} \right\rangle
\frac{T_{j2}}{\lambda_j} X_2 \right]
\left\langle K_j(\tilde{\bm x}) \right\rangle \nonumber\\
&&=-\langle K_j(\tilde{\bm x})\rangle
\left\{ {\bm g}_j  {\rm diag}(T_j) [X_1,X_2]^{\rm T}\right\}/{\lambda_j}
\end{eqnarray}

Similarly, we can modify $\left\langle\Gamma_j (\tilde {\bm n},\tilde {\bm x},t) ({\tilde n}_s- \langle{\tilde n}_s\rangle) \right\rangle$ in (\ref{eq_24}) and (\ref{eq_25}) by introducing
\begin{eqnarray}\label{eq_C16}
\langle\Gamma_j (\tilde {\bm n},\tilde {\bm x},t) ({\tilde n}_s- \langle{\tilde n}_s\rangle)\rangle
&&- \langle K_j (\tilde {\bm x})\rangle \langle \Gamma_j (\tilde {\bm n},t) ({\tilde n}_s- \langle{\tilde n}_s\rangle)\rangle
= {\rm Cov}\left[ K_j (\tilde {\bm x}), \Gamma_j (\tilde {\bm n},t) {\tilde n}_s\right] \nonumber\\
&&\cong \left\langle \frac{\partial \Gamma_j (\tilde {\bm n},t){\tilde n}_s}
{\partial {\tilde n}_1} \right\rangle
\left\langle \frac{\partial K_j(\tilde{\bm x})}{\partial {\tilde x}_1} \right\rangle X_1
+ \left\langle \frac{\partial \Gamma_j (\tilde {\bm n},t){\tilde n}_s}
{\partial {\tilde n}_2} \right\rangle
\left\langle \frac{\partial K_j(\tilde{\bm x})}{\partial {\tilde x}_2} \right\rangle X_2 \nonumber\\
&&\cong \left\langle \Gamma_j (\tilde {\bm n},t) \right\rangle
\left\langle \frac{\partial K_j(\tilde{\bm x})}{\partial {\tilde x}_s} \right\rangle X_s \nonumber\\
&&=-\frac{T_{j2}}{\lambda_j} \langle \Gamma_j (\tilde {\bm n},t) \rangle
\langle K_j(\tilde{\bm x})\rangle X_s
\end{eqnarray}

\section{Moments of Electron Number}\label{App_B}

Assume $\tilde{n}_1$, $\tilde{n}_2$ have bivariate normal distribution (from the central limit theorem under weak dependence mentioned before) with variance $D_s (t)=\langle\tilde{n}_s^2\rangle-\langle\tilde{n}_s \rangle^2$, $s=1,2$, and covariance $\Sigma(t)=\langle\tilde{n}_1 \tilde{n}_2\rangle-\langle\tilde{n}_1\rangle\langle\tilde{n}_2\rangle$. Define the moment
\begin{equation}\label{eq_H1}
M_{l_1 l_2} \equiv \langle(\tilde{n}_1-\langle\tilde{n}_1\rangle)^{l_1} (\tilde{n}_2-\langle\tilde{n}_2\rangle)^{l_2}\rangle,
\end{equation}

\noindent According to the Isserlis' theorem, the moment is equal to summing over all distinct ways of partitioning $l_1 (n_1-\langle\tilde{n}_1\rangle)$ and $l_2 (n_2-\langle\tilde{n}_2\rangle)$ into pairs.

If $l_1+l_2$ is odd, $M_{l_1 l_2}=0$.

If $l_1$, $l_2$ are both even,
\begin{eqnarray}\label{eq_H2}
M_{l_1 L_2} &=& \!\sum_{k=0}^{l_{\rm m}}
\left[(2k)!D_{12}^{2k}\right] \nonumber\\
&&\cdot\left[
\left( \begin{array}{c}
l_1\\
2k
\end{array} \right)
\left( \begin{array}{c}
l_1-2k\\
2
\end{array} \right)
\left( \begin{array}{c}
l_1-2k-2\\
2
\end{array} \right)
\cdots
\left( \begin{array}{c}
2\\
2
\end{array} \right)
\frac{D_1^{l_1/2-k}}{(l_1/2-k)!}\right] \nonumber\\
&&\cdot\left[
\left( \begin{array}{c}
l_2\\
2k
\end{array} \right)
\left( \begin{array}{c}
l_2-2k\\
2
\end{array} \right)
\left( \begin{array}{c}
l_2-2k-2\\
2
\end{array} \right)
\cdots
\left( \begin{array}{c}
2\\
2
\end{array} \right)
\frac{D_{22}^{l_2/2-k}}{(l_2/2-k)!}\right] \nonumber\\
&=& \!\sum_{k=0}^{l_{\rm m}}
\left( \begin{array}{cc}
l_1 \\
2k
\end{array} \right)
\frac{(l_1-2k)!D_{11}^{\frac{l_1}{2} - k}} {2^{k l_1/2}(l_1/2-k)!}
\left( \begin{array}{cc}
l_2 \\
2k
\end{array} \right)
\frac{(l_2-2k)!D_{22}^{\frac{l_2}{2} - k}} {2^{k l_2/2}(l_2/2-k)!}
(2k)!D_{12}^{2k} \nonumber\\
&=&\displaystyle{\!\sum\limits_{k = 0}^{l_{\rm m}}}{\frac{l_1!\,l_2! D_{11}^{\frac{l_1}{2} - k} D_{22}^{\frac{l_2}{2} - k}}{(l_1 - 2k)!!(l_2 - 2k)!!}\frac{D_{12}^{2k}}{(2k)!}},
\end{eqnarray}
\noindent where $l_{\rm{m}}$ denotes the integer part of $\min(l_1, l_2)/2$.

If $l_1$, $l_2$ are both odd,
\begin{eqnarray}\label{eq_H3}
M_{l_1 l_2} &=& \!\
\!\sum_{k=0}^{l_{\rm m}}
[(2k+1)!D_{12}^{2k+1}] \nonumber\\
&&\cdot\left[
\left( \begin{array}{c}
l_1\\
2k+1
\end{array} \right)
\left( \begin{array}{c}
l_1-2k-1\\
2
\end{array} \right)
\left( \begin{array}{c}
l_1-2k-3\\
2
\end{array} \right)\right.
\cdots\left.
\left( \begin{array}{c}
2\\
2
\end{array} \right)
\frac{D_1^{(l_1-1)/2-k}}{((l_1-1)/2-k)!}\right] \nonumber\\
&&\cdot\left[
\left( \begin{array}{c}
l_2\\
2k+1
\end{array} \right)
\left( \begin{array}{c}
l_2-2k-1\\
2
\end{array} \right)
\left( \begin{array}{c}
l_2-2k-3\\
2
\end{array} \right)\right.
\cdots\left.
\left( \begin{array}{c}
2\\
2
\end{array} \right)
\frac{D_{12}^{(l_2-1)/2-k}}{((l_2-1)/2-k)!}\right] \nonumber\\
&=& \!\sum_{k=0}^{l_{\rm m}}
\left( \begin{array}{cc}
l_1 \\
2k+1
\end{array} \right)
\frac{(l_1-2k-1)!D_{11}^{\frac{l_1-1}{2} - k}} {2^{k(l_1-1)/2}((l_1-1)/2-k)!}
\cdot\left( \begin{array}{cc}
l_2 \\
2k+1
\end{array} \right)
\frac{(l_2-2k-1)!D_{22}^{\frac{l_2-1}{2} - k}} {2^{k(l_2-1)/2}((l_2-1)/2-k)!}
(2k+1)!D_{12}^{2k+1}\nonumber\\
&=&\displaystyle{\!\sum\limits_{k = 0}^{l_{\rm m}}}{\frac{l_1!\,l_2! D_{11}^{\frac{l_1-1}{2} - k} D_{22}^{\frac{l_2-1}{2} - k}}{(l_1-1-2k)!!(l_2-1-2k)!!}\frac{D_{12}^{2k+1}}{(2k+1)!}}.
\end{eqnarray}

Finally, we have the expression of $M_{l_1 l_2}$ in (\ref{eq_22}).

\end{appendix}

\end{document}